\documentclass[12pt, a4paper]{article}
\usepackage{cmap}
\usepackage[T2A]{fontenc}		
\usepackage[utf8]{inputenc}		
\usepackage[ukrainian, english]{babel}	
\usepackage[colorlinks,hypertexnames=false]{hyperref}
\usepackage{caption}
\usepackage{cite}
\usepackage{graphicx}  
\graphicspath{{images/}{images2/}}  
\usepackage[justification=centering,singlelinecheck=false]{caption}

\usepackage{amssymb,amsfonts,amsmath,amsthm}

\usepackage[usenames]{color}
\usepackage{colortbl}
\usepackage[dvipsnames]{xcolor}

\usepackage{array,tabularx,tabulary,booktabs} 

\renewcommand{\phi}{\varphi}

\counterwithin{equation}{section}
\usepackage{indentfirst}

\usepackage[titletoc, title]{appendix}

\begin{document}
\title{Effective interaction of Chern-Simons boson with fermions}
\author{Ivan Hrynchak${}^1$, Oleksandr Khasai${}^2$, Yuliia  Borysenkova${}^{3}$,\\
Mariia Tsarenkova${}^1$, Volodymyr Gorkavenko${}^{1,2}$\thanks{Corresponding author. E-mail address: \textit{gorkavol@knu.ua} (V. Gorkavenko).} \vspace{1em}\\
${}^1$ \it \small Faculty of Physics, Taras Shevchenko National University of Kyiv,\\ 
\it \small 64, Volodymyrs'ka str., Kyiv 01601, Ukraine,\\
${}^2$ \it \small Bogolyubov Institute for Theoretical Physics, National Academy of Sciences of Ukraine,\\
\it \small 14-b Metrolohichna str., Kyiv 03143, Ukraine,\\
${}^3$ \it \small Institut de Física d'Altes Energies (IFAE), The Barcelona Institute of Science and Technology (BIST),\\
\it \small Campus UAB, 08193 Bellaterra (Barcelona), Spain}

\date{}

\maketitle

\abstract{We consider a vector extension of the Standard Model (SM) with a Chern–Simons–type interaction. This extension introduces a new massive vector boson (the Chern–Simons (CS) boson) that does not couple directly to SM fermions at tree level. We analyze the effective loop-induced interaction of this new vector boson with SM fermions and study its renormalizability in the $R_\xi$ gauge.
We find that, in the effective interaction between the CS boson and same-flavor fermions, the divergent contributions from individual loop diagrams do not cancel when all relevant diagrams are taken into account. In contrast, for interactions involving fermions of different flavors, the corresponding loop-induced contributions are finite and well defined. This indicates that, in the low-energy limit, the theory exhibits nonrenormalizable behavior in the sector describing the loop-induced interaction of the CS boson with same-flavor fermions. 
The interaction terms between the CS boson and same-flavor fermions, characterized by divergent coefficients, are identified and must be treated within the framework of effective field theory. Finally, we derive the leading-order effective Lagrangian describing the interaction of a GeV-scale CS boson with SM fermions and discuss the number of independent parameters entering this Lagrangian. The leading-order interaction we obtained turns out to be similar to the interaction of a $Z^\prime$ boson with SM fermions.}

\section{Introduction}\label{sec:intro}

The motivation for researching various SM extensions is that the SM \cite{Cottingham:2007zz} is an incomplete theory, as it fails to provide an explanation for such fundamental phenomena as active neutrino oscillations (see, e.g., \cite{Bilenky:1987ty,Strumia:2006db,deSalas:2017kay}), the baryon asymmetry of the Universe (see, e.g., \cite{Steigman:1976ev,Riotto:1999yt,Canetti:2012zc}), and the existence of dark matter (see, e.g., \cite{Peebles:2013hla,Lukovic:2014vma,Bertone:2016nfn}).  These problems indicate that there may be a hidden sector in the SM with new particles, some of which may not be directly relevant to resolving  SM problems. 

These particles remain undetected either because they are too massive to be produced at present accelerators or because they interact only feebly with SM particles. They cannot be produced in the present experiments if they are sufficiently heavy. On the other hand, if they are light, they could potentially be observed at current and upcoming intensity frontier experiments \cite{Beacham:2020,Lanfranchi:2020crw} such as MATHUSLA \cite{Curtin:2018mvb}, FACET \cite{Cerci:2021nlb}, FASER \cite{FASER:2018ceo,FASER:2018eoc}, SHiP \cite{Anelli:2015pba, Alekhin:2015byh}, DUNE \cite{DUNE:2015lol,DUNE:2020fgq}, LHCb \cite{Chen:2021ftn,Gorkavenko:2023nbk}, and others.  

At the same time, the nature of the new physics particles remains unknown. They could be scalars \cite{Patt:2006fw,Bezrukov:2009yw,Boiarska:2019jym}, pseudoscalars (axion-like particles) \cite{Peccei:1977hh, Weinberg:1977ma, Wilczek:1977pj,Choi:2020rgn}, fermions \cite{Asaka:2005pn,Asaka:2005an,Bondarenko:2018ptm,Boyarsky:2018tvu}, or vectors (dark photons or $Z^\prime$ bosons) \cite{Okun:1982xi,Holdom:1985ag,Fabbrichesi:2020wbt,Langacker:2008yv}. For comprehensive discussions, see, for example, the reviews in \cite{Curtin:2018mvb,Alekhin:2015byh}.

In this paper, we consider one of the possible extensions of the Standard Model (SM), namely the vector extension with a Chern-Simons type
interaction. In this extension, there is a new massive vector boson, which we will call the Chern-Simons (CS) boson in the following. 
The Chern-Simons type interaction we are considering arises in different theoretical models, including extra dimensions and string theory, 
see e.g. \cite{Antoniadis:2000ena,Coriano:2005own,Anastasopoulos:2006cz,Harvey:2007ca,Anastasopoulos:2008jt,Kumar:2007zza}.

The minimal gauge-invariant Lagrangian of the interaction of the CS bosons with SM particles takes the form 
of dimension-six operators and is presented in  \cite{Alekhin:2015byh,Antoniadis:2009ze}, where the SM is extended to a theory with symmetry $SU_C(3)\times SU_W(2)\times U_Y (1) \times U_X(1)$ \vspace{-0.5em}
\begin{align}
    \mathcal{L}_1&=\frac{C_Y}{\Lambda_Y^2}\cdot X_\mu (\mathfrak D_\nu H)^\dagger H B_{\lambda\rho} \cdot\epsilon^{\mu\nu\lambda\rho}+h.c.,\label{L1} \\
    \mathcal{L}_2&=\frac{C_{SU(2)}}{\Lambda_{SU(2)}^2}\cdot X_\mu (\mathfrak D_\nu H)^\dagger F_{\lambda\rho} H\cdot\epsilon^{\mu\nu\lambda\rho}+h.c.,\label{L2}  
\end{align}
where $\Lambda_Y$, $\Lambda_{SU(2)}$ are new scales of the theory, $C_Y$ and $C_{SU(2)}$ are new dimensionless coupling constants, and $\epsilon^{\mu\nu\lambda\rho}$ is the Levi–Civita symbol ($\epsilon^{0123}=+1$). The field $X_\mu$ denotes the CS vector boson with mass $M_X$, and $H$ is the Higgs doublet.  The field strength tensors are defined as $B_{\mu\nu}=\partial_\mu B_\nu-\partial_\nu B_\mu$ and $F_{\mu\nu}=-ig\sum\limits_{i=1}^3 \frac{\tau^i}{2}V^i_{\mu\nu}$ for the $U_Y(1)$ and $SU_W(2)$ gauge fields of the SM, respectively. The gauge invariance of the Lagrangians \eqref{L1} and \eqref{L2} is ensured by the fact that $X_\mu$ is a Stueckelberg field \cite{Ruegg:2003ps,Kribs:2022gri}.

After electroweak symmetry breaking, Lagrangians \eqref{L1}, \eqref{L2}  generate  (among other terms
of higher dimensions) Lagrangians of three-field interactions of the CS vector boson with vector fields of the SM and the Higgs boson. In the unitary gauge, they have forms\vspace{-0.5em}
\begin{equation}\label{Lcs}  
     \mathcal{L}^{(4)}_{CS}=
     c_z \epsilon^{\mu\nu\lambda\rho} X_\mu Z_\nu \partial_\lambda Z_\rho +c_\gamma \epsilon^{\mu\nu\lambda\rho} X_\mu Z_\nu \partial_\lambda A_\rho+\left\{ c_w \epsilon^{\mu\nu\lambda\rho} X_\mu W_\nu^- \partial_\lambda W_\rho^+ + h.c.\right\},
\end{equation}
\begin{equation}\label{Lcs5}  
     \mathcal{L}^{(5)}_{CS}=c_{\gamma h}\epsilon^{\mu\nu\lambda\rho} X_\mu \frac{\partial_\nu h}v \partial_\lambda A_\rho +
     c_{z h}\epsilon^{\mu\nu\lambda\rho} X_\mu \frac{\partial_\nu h}v \partial_\lambda Z_\rho,
\end{equation}
where \eqref{Lcs} contains dimension-four operators, but in \eqref{Lcs5} we have dimension-five operators. 
Here  $h$ is the Higgs boson field and $v$ is its vacuum expectation value; $A_\mu$ is the electromagnetic field; $W^\pm_\mu$ and $Z_\mu$ are fields of the weak interaction. Dimensionless coefficients $c_{\gamma h}$, $c_{z h}$, $c_z$ and $c_\gamma$ are real, but $c_w$  can be complex.

As evident from Eqs.\eqref{Lcs} and \eqref{Lcs5}, the CS vector boson $X_\mu$ does not couple directly to SM fermions.   {This is a defining feature of the SM extension considered in this work, in which direct couplings between $X_\mu$ and SM fermions are forbidden by construction.} In \cite{Antoniadis:2009ze}, this was achieved by imposing that all SM fermions are neutral under the $U_X(1)$ gauge group.   {Instead, the interaction between $X_\mu$ and the SM particles is generated indirectly through heavy chiral fermions charged under both $U_X(1)$ and the electroweak gauge group. Once these heavy fermions are integrated out, their nontrivial anomaly-cancellation structure induces D'Hoker--Farhi-type (generalized Chern--Simons) interactions in the low-energy effective theory. In our framework, such anomaly-induced effects are described by the dimension-six operators of Eqs.(\ref{L1}) and (\ref{L2}).}

The original motivation   {for constructing a theory without direct couplings between the new vector boson and SM fermions} can be summarized as follows. If SM fermions carried nonzero $U_X(1)$ charges, the corresponding gauge boson would generically appear as a resonance in $pp$ collisions through processes of the type $q\bar q\rightarrow X \rightarrow f\bar f$, leading to phenomenology closely resembling that of a conventional $Z^\prime$
boson \cite{Langacker:2008yv}. To avoid, or at least strongly suppress, such resonant production channels, the authors of \cite{Antoniadis:2009ze} therefore proposed to exclude direct couplings between the CS boson and SM fermions.

 If one rewrites the coefficients before operators in \eqref{L1}, \eqref{L2}  as
 $C_Y/\Lambda_Y^2=C_1/v^2$ and $C_{SU(2)}/\Lambda_{SU(2)}^2=C_2/v^2$, where $C_1=c_1+ i \tilde c_{1}$ and  $C_2=c_2+ i \tilde c_{2}$ are dimensionless coefficients, and compares terms of three-field interactions in Lagrangians \eqref{L1}, \eqref{L2} and Lagrangians  \eqref{Lcs}, \eqref{Lcs5}, then one can obtain\vspace{-0.5em}
\begin{align}
&    c_{\gamma h} = 2 c_1 \cos\theta_W - g \tilde c_2 \sin\theta_W, \label{cgammah} \\
&    c_{z h} = -2 c_1 \sin\theta_W - g \tilde c_2 \cos\theta_W, \label{czh} \\
&    c_z=  -\tilde c_1g' + \frac{c_{2}}2 g^2, \label{L7}\\
&    c_\gamma = +\tilde c_1 g + \frac{c_{2}}2 g g^{'} , \label{L8}\\
&    c_w= \frac{c_2+i \tilde c_2}2 g^2\equiv \Theta_{W1}+i \Theta_{W2},\label{L9}
\end{align}
where $e=g \sin \theta_W=g' \cos \theta_W$ is the absolute value of the electron's electric charge and $\theta_W$ is the Weinberg angle. It should be noted that the contribution to the effective interaction of CS bosons with vector fields of SM can also come from higher dimension operators \cite{Alekhin:2015byh}. In this case, relations \eqref{cgammah} -- \eqref{L9} should be modified. 
 
The question of the renormalizability of three-field interactions given by dimension-four operators in \eqref{Lcs} remains interesting. This question was considered in \cite{Dror:2017ehi,Dror:2017nsg,Borysenkova:2021ydf}, where the effective loop interaction of the CS bosons with quarks of different flavors $q_i q_j$ was analyzed. It was shown that, in this case, the divergent parts of the loop diagrams (containing only the interaction of the 
CS boson with $W^\pm$ bosons) are proportional to an off-diagonal element of the unit matrix $(V^+V)_{ij}$ ($V$ is the Cabibbo–Kobayashi–Maskawa matrix)   {and therefore cancel as a consequence of the Glashow--Iliopoulos--Maiani (GIM) mechanism~\cite{Glashow:1970gm,Maiani:2013fpa}. Such a cancellation is a manifestation of a generic feature of flavor-changing neutral-current amplitudes in the SM.}
It becomes obvious that the effective interaction of the CS bosons with quarks of the same flavors or leptons via loop diagrams with $W^\pm$ bosons will contain divergence \cite{Gorkavenko:2024ivy}. This divergence cannot be removed via counterterms of the CS boson interaction with fermions because the initial Lagrangian \eqref{Lcs} does not include these terms. The question of loop interaction between the CS bosons and fermions of the same flavors is extremely important. Without resolving it, we cannot even calculate such a simple reaction as $X\rightarrow \ell^- \ell^+$.

In \cite{Borysenkova:2024xno}, the problem of effective loop interaction of the CS boson with leptons was considered by taking into account all possible diagrams in the unitary gauge using only dimension-four operators of the Lagrangian \eqref{Lcs}. It was shown that loop divergences could not be eliminated.

In this paper, we consider the effective loop interaction of the CS bosons with fermions of the same flavors in the general case of $R_\xi$ gauge and check whether there will be a cancellation of divergences when accounting for all corresponding diagrams. 
 Since, in the non-unitary gauge, the interactions of the Chern–Simons boson with Goldstone bosons are described by dimension-five operators, our calculations will also include its interactions with Higgs bosons, appearing in the Lagrangian \eqref{Lcs5}, in the form of dimension-five operators as well.

Our motivation for considering the problem in the general case of a non-unitary gauge is driven by the requirement of consistency between calculations performed in the unitary and $R_\xi$ gauges; in particular, obtaining identical results for divergent contributions in different gauges would serve as an important check of the correctness of our computations, which is a non-trivial issue for non-Abelian field theories. As in \cite{Borysenkova:2024xno}, we will consider the effective interaction of the CS bosons only with leptons to avoid unnecessary complications from the CKM matrix elements.

Having analyzed the renormalizability of the loop-induced interaction between the CS boson and SM fermions, we then construct the corresponding effective Lagrangian describing their interaction.

Determining the interaction of the CS boson with SM fermions will allow us to identify its dominant production and decay channels, thereby providing the basis for future studies of the sensitivity of intensity-frontier experiments to the CS boson, see, e.g., \cite{Ovchynnikov:2023cry}. In this context, it is important to emphasize that the CS boson was included among the physics targets of the SHiP experiment proposal \cite{Alekhin:2015byh}.

The structure of the paper is as follows. Section 2 is devoted to the analysis of the interaction of the CS boson in the $R_\xi$
 gauge. Sections 3 and 4 address the calculation of divergent contributions in loop diagrams describing the interaction of the CS boson with SM fermions of the same flavor. In Section 5 we demonstrate that these divergent contributions do not cancel even after taking into account all relevant diagrams. Section 6 is dedicated to constructing an effective Lagrangian for the interaction of the GeV-scale CS boson with SM fermions and to discussing the number of independent parameters appearing in this Lagrangian. The summary and final discussion are presented in Section 7. Technical details and calculations are given in the Appendices.

\section{Interactions of the CS bosons in the $R_\xi$ gauge}

In the non-unitary gauge, the doublet of the Higgs field obtains the form,
\begin{equation}\label{Hgold}
    H = \begin{pmatrix}
    \phi^+\vspace{0.5em}\\
    \dfrac{v + h + i\phi_z}{\sqrt2} \end{pmatrix},
\end{equation}
where $\phi^+$, $\phi_z$ are charged and neutral Goldstone bosons, and $h$ is the Higgs field. In this case, the simple three-particle interaction of the CS boson, as described by the Lagrangians \eqref{Lcs5} and \eqref{Lcs}, is supplemented by interactions with Goldstone bosons, whose explicit forms are derived from the Lagrangians \eqref{L1} and \eqref{L2}:
\begin{align}
    & \mathcal{L}^{X\phi^\mp W^\pm}  = \frac{2i}{gv}\left( c_w X_{\mu} \partial _{\nu} \phi^{-} \,\partial_{\lambda} W_{\rho}^{+} - c_w^{*} X_{\mu} \partial_{\nu} \phi^{+} \,\partial_{\lambda} W_{\rho}^{-}  \right)\epsilon^{\mu\nu\lambda\rho},\\
    & \mathcal{L}^{X\phi_z Z} =  \frac{c_z g}{2\cos\theta_W} X_{\mu} \frac{\partial_{\nu} \phi_{Z}}{v} \partial_{\lambda} Z_{\rho} \epsilon^{\mu\nu\lambda\rho},\\
     & \mathcal{L}^{X\phi_z A}= \frac{c_{\gamma} g}{2\cos\theta_W} X_{\mu} \frac{\partial_{\nu} \phi_{Z}}{v} \partial_{\lambda} A_{\rho}\epsilon^{\mu\nu\lambda\rho}.
\end{align}
Using the rules of Feynman diagrams, see Appendix \ref{AppA}, we calculate the process of the decay of the massive CS boson into a charged lepton pair $\ell^+\ell^-$ ($\ell=e,\mu,\tau$) in $R_\xi$ gauge. The results for decay into neutrinos can be easily obtained by a simple change of notations.

\section{Divergences in the triangle diagrams}\label{sec:triadagr}

\subsection{Vertices $XWW$ and $X\phi^\mp W^\pm$}

\begin{figure}[h!]\centering
            \label{fig:eevWphiX}
            \includegraphics[width = 0.85\textwidth]{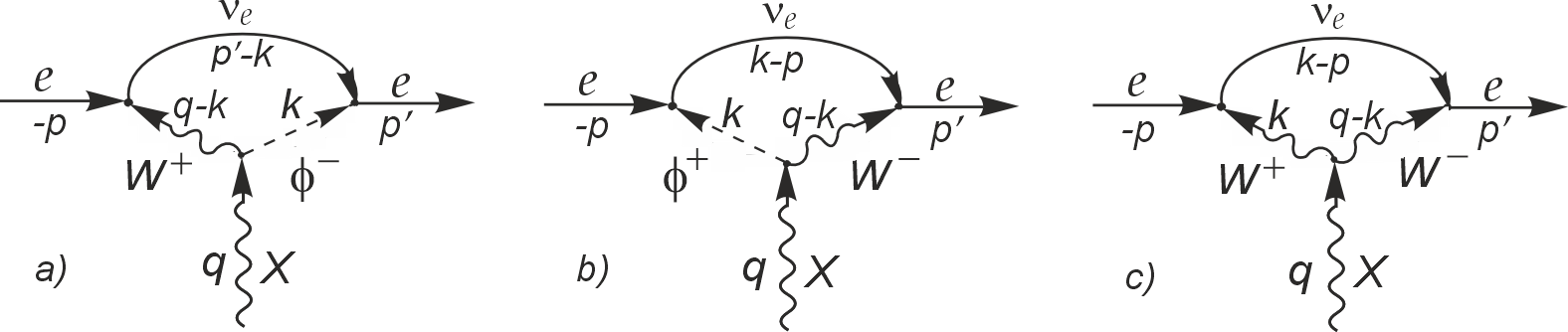}
            \captionsetup{width=0.9\textwidth}
            \caption{Triangle diagrams for lepton production via CS-boson decay, mediated by either the interaction with two \(W\) bosons or with a \(W\) boson and a charged \(\phi\) boson.}
            \label{fig:Rxi_triangle_diag}
        \end{figure}

Triangle diagrams for lepton production via CS-boson decay, mediated either by the interaction of the CS boson with two \(W\) bosons or with a \(W\) boson and a charged \(\phi\) boson, are shown in Fig.\ref{fig:Rxi_triangle_diag}.

The amplitude of the process for the first two diagrams of Fig.\ref{fig:Rxi_triangle_diag} is given by
 \begin{multline}\label{XphiW}
     M_{fi}^{X\phi W} = -
  \dfrac{2 m_\ell}{v^2} \bar{\ell}(p^\prime) \left\{c_w
           \gamma^\alpha \gamma_\rho \hat{P}_L       [I_{\nu\alpha}^{ W}(p',\xi_W)-I_{\nu}^{W}(p',\xi_W)p'_{\alpha}]+\right. \\ \left.+ c_w^* \gamma_\rho \hat{P}_L \gamma^\alpha     [I_{\nu\alpha}^{ W}(p,\xi_W)-I_{\nu}^{ W}(p,\xi_W)p_{\alpha}] \right\}\ell(-p)  q_\lambda \varepsilon^{\lambda_X}_\mu(q)\varepsilon^{\mu \nu \lambda \rho}
    \end{multline}
and the amplitude of the process for the third diagram of Fig.\ref{fig:Rxi_triangle_diag} is given by (see details in Appendices \ref{AppB} and \ref{AppC})
\begin{multline}\label{XWW}
        M_{fi}^{XWW}=      \dfrac{g^2}{2} \,\bar{\ell}(p^\prime) \gamma^\delta \gamma^\alpha \gamma^\gamma \hat{P}_L \ell(-p)\, \varepsilon_\mu^X(q) \varepsilon^{\mu \nu \lambda \rho} g_{\rho \gamma} g_{\delta \nu} \times \\
        \times [c_w  \{I_{\alpha\lambda}^{W}(p,1)-I_{\lambda}^{W}(p,1) p_\alpha\}-
        c_w^* \{I_{\alpha}^{W}(p,1) q_\lambda -I_{\alpha\lambda}^{W}(p,1) -  \\ -p_\alpha q_\lambda I^{W}(p,1)+  I_\lambda^{W}(p,1) p_\alpha\} + (1-\xi_W) \cdot(\mbox{c.t.}) ],
            \end{multline}
       where 
        \begin{align}
              &  I^{ W}(a,\xi_W)\!=\!\!\!\int\!\! \dfrac{d^4 k}{(2\pi)^4}
         \dfrac{1}{(m_\nu^2-(k-a)^2)(M_W^2-(k-q)^2)(\xi_W M_W^2-k^2)}= \frac{i\pi^2}{(2\pi)^4}\!\int\!  \frac{\{dx\}_3}{D^{ W}(a,\xi_W) }, \\
          &      I_{\nu}^{ W}(a,\xi_W)=\int \dfrac{d^4 k}{(2\pi)^4}
         \dfrac{k_\nu }{(m_\nu^2-(k-a)^2)(M_W^2-(k-q)^2)(\xi_W M_W^2-k^2)}= \frac{i\pi^2}{(2\pi)^4}\int \{dx\}_3 \frac{(y q-x a)_\nu}{D^{ W}(a,\xi_W)},\\
       &  I_{\nu\alpha}^{ W}(a,\xi_W)=\int \dfrac{d^4 k}{(2\pi)^4}
         \dfrac{k_\nu k_\alpha}{(m_\nu^2-(k-a)^2)(M_W^2-(k-q)^2)(\xi_W M_W^2-k^2)}= -\frac{i\pi^2}{(2\pi)^4}\frac{g_{\nu\alpha}}{4}\ln \frac{\Lambda^2}{M_W^2} +\nonumber \\&
         + \frac{i\pi^2}{(2\pi)^4}\int \{dx\}_3 \frac{(y q-x a)_\nu (y q-x a)_\alpha}{D^{ W}(a,\xi_W)}
         + \frac{i\pi^2}{(2\pi)^4} \frac{g_{\nu\alpha}}{2}\int \{dx\}_3
         \ln\frac{D^{ W}(a,\xi_W)}{M_W^2 x}. \label{k2integral} 
\end{align} 
The following notations are used here:
\begin{align} 
        &\int \{dx\}_3f(x,y,z)=  \int\limits_0^1 dx \int\limits_0^1 dy\int\limits_0^1 dz \,\, \delta(x\!+\!y\!+\!z\!-\!1)f(x,y,z)=\int\limits_0^1 dx \int\limits_0^{1-x} dy \,f(x,y,1\!-\!x\!-\!y),\label{Intdx3}\\
        & D^{W}(a,\xi_W)=M_W^2 (y+\xi_W z)-a^2 x(1-x) -M_X^2 y(1-y)-2xy (aq) +  x m_\nu^2.\label{DphiW} 
    \end{align}  
$\hat P_{R(L)}=(1+(-)\gamma^5)/2$ are projection operators, $p$ is the 4-momentum of the lepton antiparticle, $p^\prime$ is the 4-momentum of the lepton particle,  $q=p+p^\prime$ is the 4-momentum of the CS boson, $q^2=M_X^2$, $\Lambda $ is a cut-off parameter, and the abbreviation c.t. denotes convergent terms of a more cumbersome form, which we do not write out.
    
It should be noted that we used the technique of the $\alpha$ (Schwinger) representation and Frullani-based regularization (an analogue of Pauli–Villars regularization) \cite{zuber2012quantum} in the form
    \begin{equation}
        \int\limits_0^\infty \frac{e^{-iR (D-i \varepsilon)}}{R}dR \rightarrow \int\limits_0^\infty \frac{e^{-iR (D-i \varepsilon)}-e^{-iR (D-i \varepsilon)|_{m_{virt}=\Lambda \rightarrow \infty}}}{R}dR=\ln \frac{D|_{m_{virt}=\Lambda \rightarrow \infty}}{D},
    \end{equation}
where $m_{virt}$ is the mass of the virtual fermion in the loop, see Appendix \ref{AppC}. Of course, the neutrino mass should be zero ($m_\nu=0$) in the above relations, but we will temporarily leave it to simplify comparison with further formulas. It will also be useful for analyzing CS-boson decays into neutrinos, which can be obtained by a simple change of notation.
    
As one can see, only tensor $I_{\nu\alpha}$ contains divergent terms, so
    we can write the divergent terms of amplitudes \eqref{XphiW} and \eqref{XWW} in the form
 \begin{align}
      &  M_{fi,div}^{X\phi W}=
      i  \dfrac{ m_\ell}{v^2} \Lambda_1 \,\bar{\ell}(p^\prime)  \gamma_{\lambda}\gamma_\rho [i\Theta_{W2}-\gamma^5 \Theta_{W1}] \ell(-p)   q_\nu \varepsilon^{\lambda_X}_\mu(q)\varepsilon^{\mu \nu \lambda \rho},\label{XphiW,div}\\
      &  M_{fi,div}^{XWW}=    i \Theta_{W1} \frac{g^2}{2} \Lambda_1   \,\bar{\ell}(p^\prime) \hat{P}_R \gamma_\rho \gamma_\lambda \gamma_\nu \hat{P}_L \ell(-p)\, \varepsilon_\mu^{\lambda_X}(q) \varepsilon^{\mu \nu \lambda \rho}, \label{XWW,div}
\end{align}
where, for simplicity, we used a new parameter
\begin{equation}
    \Lambda_1 = \frac{\pi^2}{2(2\pi)^4}  \ln\frac{\Lambda^2}{M_W^2}\rightarrow \infty.
\end{equation}

\subsection{Vertices $XZZ$ and $X\phi_z Z$}

\begin{figure}[h]
        \centering
        \includegraphics[width = 0.7\textwidth]{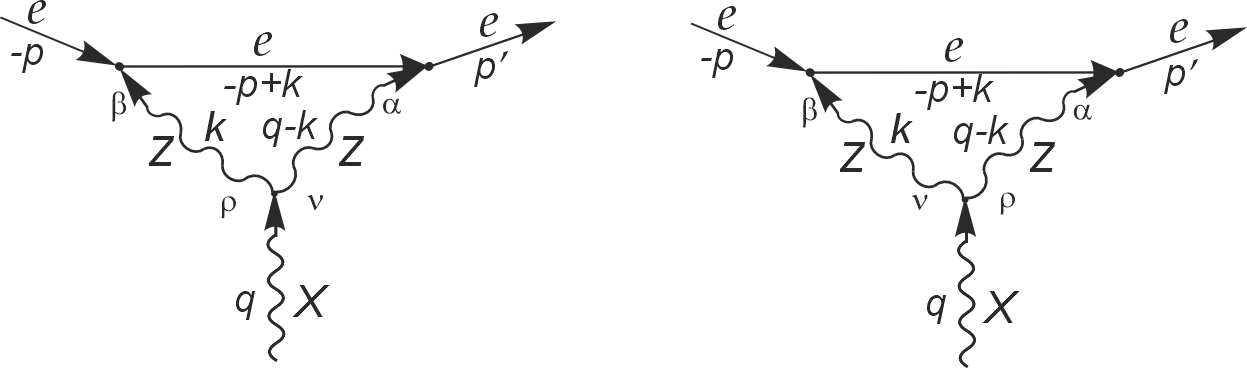} 
        \captionsetup{width=0.9\textwidth}
        \caption{Triangle diagrams for lepton production via CS-boson decay, mediated by the interaction of the CS boson with two $Z$ bosons.}
        \label{fig:XZZ_Rx0i}
    \end{figure}

\begin{figure}[h]
        \centering
        \includegraphics[width = 0.7\textwidth]{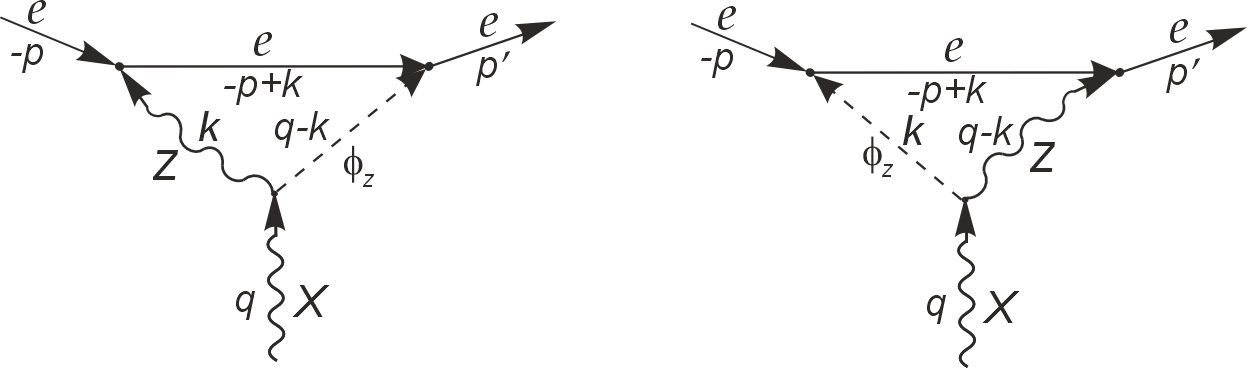} 
        \captionsetup{width=0.9\textwidth}
        \caption{Triangle diagrams for lepton production via CS-boson decay, mediated by the interaction of the CS boson with a $Z$ and a neutral $\phi_z$ boson.}
        \label{fig:XZphiz}
        \end{figure}

Triangle diagrams for lepton production via CS-boson decay, mediated either by the interaction of the CS boson with two $Z$ bosons or with a $Z$ and a neutral $\phi_z$ boson, are presented in Fig.\ref{fig:XZZ_Rx0i} and Fig.\ref{fig:XZphiz}. Two diagrams in Fig.\ref{fig:XZZ_Rx0i} arise from the interaction of the CS boson with two $Z$ bosons \eqref{Lcs}, corresponding to two possible positions for the $Z$ boson's line with a derivative.

The amplitude of the process for diagrams in Fig.\ref{fig:XZZ_Rx0i} is given by
     \begin{multline}
            M_{fi}^{XZZ} = \frac{g^2 c_z}{4\cos^2\theta_W} \bar\ell(p^\prime)\, \hat{\overline{P}}_Z \gamma^\alpha \gamma^\delta \gamma^\beta \hat P_Z \ell(-p) \epsilon^{\lambda_X}_\mu(q)\, \epsilon^{\mu\nu\lambda\rho} \, \left[  g_{\alpha\rho} g_{\beta\nu} (-I^{Z}_{\delta\lambda}(0,q;1) + I^{Z}_{\lambda}(0,q;1) p_\delta) + \right. \\  +  g_{\alpha\rho} g_{\beta\nu} (-I^{Z}_{\delta\lambda}(0,q;1) + I^{Z}_{\lambda}(0,q;1) p_\delta + I^{Z}_{\delta}(0,q;1) q_\lambda - I^{Z}(0,q;1) p_\delta q_\lambda) +\\ \left.+(1-\xi_Z) \cdot (\rm c.t.) \vphantom{I^{ZZ}}\right],
        \end{multline}
   where
        \begin{align}
         &  I^{Z}(a,b;\xi_Z) = \int \dfrac{d^4 k}{(2\pi)^4}
         \dfrac{1}{(m_\ell^2-(k-p)^2)(M_Z^2-(k-a)^2)(\xi_Z M_Z^2-(k-b)^2)}= \nonumber \\
         & \hspace{25em} = \frac{i\pi^2}{(2\pi)^4}\!\!\int \{dx\}_3 \frac{1}{D^Z (a,b;\xi_Z) }, \\
         &  I^{Z}_{\lambda}(a,b;\xi_Z) = \int \dfrac{d^4 k}{(2\pi)^4}
         \dfrac{k_\lambda }{(m_\ell^2-(k-p)^2)(M_Z^2-(k-a)^2)(\xi_Z M_Z^2-(k-b)^2)}= \nonumber \\
         & \hspace{25em} = \frac{i\pi^2}{(2\pi)^4}\int \{dx\}_3 \frac{(xp+yq)_\lambda}{D^Z (a,b;\xi_Z)},
         \end{align}
         \begin{align}
         &  I^{Z}_{\delta\lambda}(a,b;\xi_Z) = \int \dfrac{d^4 k}{(2\pi)^4}
         \dfrac{k_\delta k_\lambda}{(m_\ell^2-(k-p)^2)(M_Z^2-(k-a)^2)(\xi_Z M_Z^2-(k-b)^2)}=-\frac{i\pi^2}{(2\pi)^4}\frac{g_{\delta\lambda}}{4}\ln \frac{\Lambda^2}{M_W^2} \nonumber \\&  
        \hspace{4em} + \frac{i\pi^2}{(2\pi)^4}\int \{dx\}_3 \frac{(xp+yq)_\delta (xp+yq)_\lambda}{D^Z (a,b;\xi_Z)}
         + \frac{i\pi^2}{(2\pi)^4} \frac{g_{\delta\lambda}}{2}\int \{dx\}_3
         \ln\frac{D^Z (a,b;\xi_Z)}{M_W^2 x}, \\
         &  D^Z (a,b;\xi_Z) =(xp+ya+zb)^2 + y(M_Z^2-a^2) + z(\xi_Z M_Z^2-b^2)
        \end{align} 
        and
        \begin{equation}\label{PSdec}
    \hat P_Z=t_3^\ell (1-\gamma^5)-2q_\ell \sin^2\theta_W,\qquad  \hat{\overline{P}}_Z =t_3^\ell (1+\gamma^5)-2q_\ell \sin^2\theta_W,
\end{equation}
$q_\ell$ is the electric charge of the lepton $\ell$ in units of proton charge, and $t_3^{\ell}$   is the third component of the weak isospin ($+1/2$ for neutrinos and $-1/2$ for electrically charged leptons). 

        The divergent part of this amplitude is
        \begin{equation}
            M_{fi,div}^{XZZ} =ic_z \frac{g^2 }{4\cos^2\theta_W} \Lambda_1 \, \bar\ell(p^\prime)\,\, \hat{\overline{P}}_Z (\gamma_\rho \gamma_\lambda \gamma_\nu) \epsilon^{\mu\nu\lambda\rho} \hat P_Z \ell(-p) \epsilon^{\lambda_X}_\mu(q).
        \end{equation}

The amplitude of the process for the diagrams in Fig.\ref{fig:XZphiz} is given by
 \begin{multline}
             M_{fi}^{X\phi_z Z} = - \frac{ m_\ell}{v^2} c_z \bar\ell (p') \left[ \gamma^5 \left( I^{Z}_\lambda (0,q;\xi_Z) \cdot (m_\ell - \not{p}) q_\nu + I^{Z}_{i\lambda} (0,q;\xi_Z) \cdot \gamma^i q_\nu \right) \gamma_\rho \hat P_Z + \right. \\ \left. + \gamma_\rho \hat P_Z \left( I^{Z}_\nu (q,0;\xi_Z) \cdot (m_\ell - \not{p}) q_\lambda + I^{Z}_{i\nu} (q,0;\xi_Z) \cdot \gamma^i q_\lambda \right) \gamma^5 \right] \ell (-p) \epsilon^{\lambda_X}_\mu (q) \epsilon^{\mu\nu\lambda\rho},
        \end{multline}
        The divergent part of this amplitude is
        \begin{equation}
          M_{fi,div}^{XZ\phi_z} =  ic_z\frac{ m_\ell  (t_3^\ell-2q_\ell \sin^2\theta_W)}{ v^2}\Lambda_1 \, \bar\ell(p^\prime)  \gamma^5  q_\nu \gamma_\lambda \gamma_\rho   \ell(-p) \epsilon^{\lambda_X}_\mu(q)\, \epsilon^{\mu\nu\lambda\rho}.
        \end{equation}

\subsection{Vertices $XZA$ and $X\phi_z A$}

\begin{figure}[h]
        \centering
        \includegraphics[width=0.7\textwidth]{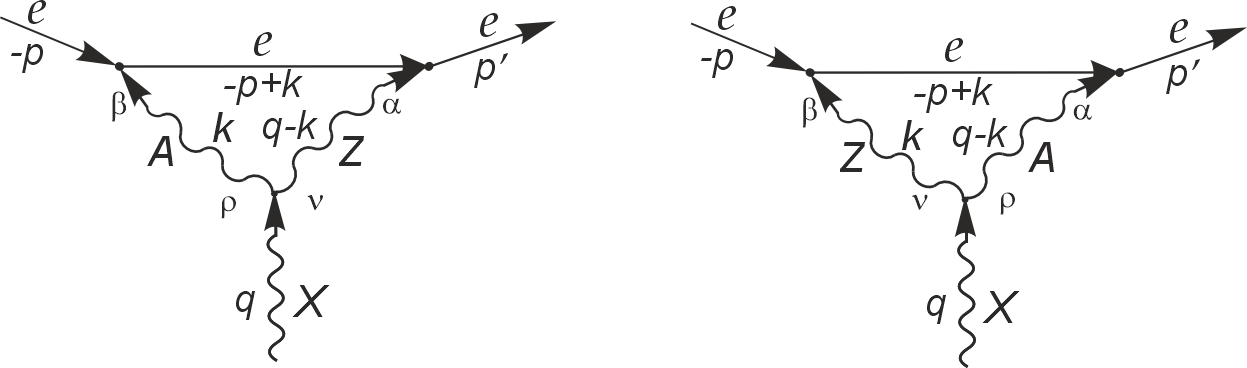} 
        \captionsetup{width=0.9\textwidth}
        \caption{Triangle diagrams for lepton production via CS-boson decay, mediated by the interactions of the CS boson with photons and \(Z\) bosons.}
        \label{fig:XZA_Rxi}
    \end{figure}

\begin{figure}[h]
        \centering
        \includegraphics[width=0.7\textwidth]{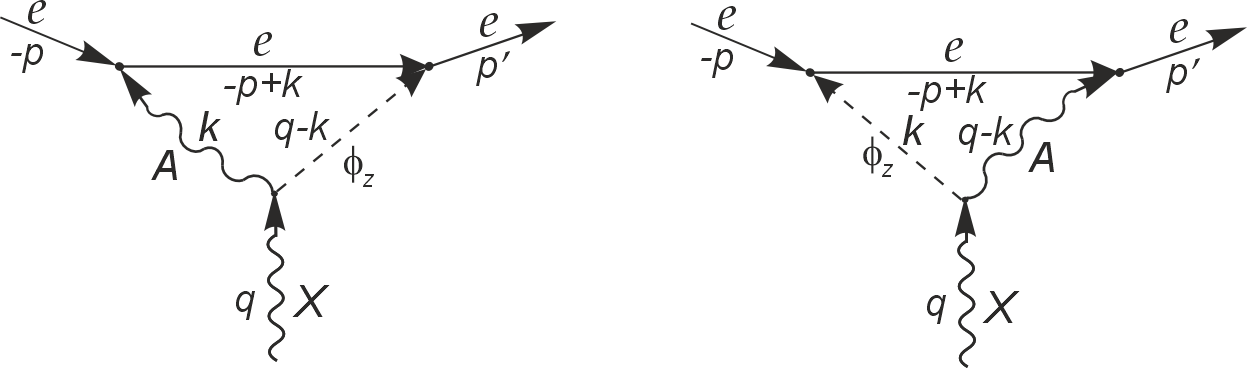} 
        \captionsetup{width=0.9\textwidth}
        \caption{Triangle diagrams for lepton production via CS-boson decay, mediated by the interactions of the CS boson with photons and neutral \(\phi_Z\) bosons.}
        \label{fig:XAphiz}
        \end{figure}

Triangle diagrams for lepton production via CS-boson decay, mediated by the interactions of the CS boson with photons and \(Z\) bosons, or with photons and neutral \(\phi_Z\) bosons, are shown in Figs.~\ref{fig:XZA_Rxi} and~\ref{fig:XAphiz}.

The amplitude of the process for the diagrams in Fig.\ref{fig:XZA_Rxi} is given by
   \begin{multline}
             M_{fi}^{XZA} = -\frac{e q_\ell g}{2\cos\theta_W} c_\gamma \times \\ \bar\ell(p^\prime)  \left[ \hat{\overline{P}}_Z \gamma^\alpha \left( I^{A}_\lambda (q,0;1) g_{\nu\alpha} g_{\rho\beta} (m_\ell - \not{p}) + I^{A}_{\delta\lambda} (q,0;1)  g_{\nu\alpha} g_{\rho\beta} \gamma^\delta - (1-\xi_Z) \cdot (\rm c.t.) \right) \gamma^\beta + \right. \\ \left. + \gamma^\alpha \left( -I^{A}_\lambda (0,q;1) g_{\rho\alpha} g_{\nu\beta} (m_\ell - \not{p}) + I^{A} (0,q;1)  g_{\rho\alpha} g_{\nu\beta} q_\lambda (m_\ell - \not{p}) + I^{A}_\delta (0,q;1) \cdot \gamma^\delta q_\lambda g_{\rho\alpha} g_{\nu\beta} - \right. \right. \\ \left. \left. - I^{A}_{\delta\lambda} (0,q;1)  \gamma^\delta g_{\rho\alpha} g_{\nu\beta} + (1-\xi_Z)\cdot (\rm c.t.) \right) \gamma^\beta \hat P_Z \right] \ell(-p) \epsilon^{\lambda_X}_\mu(q)\, \epsilon^{\mu\nu\lambda\rho},
        \end{multline}
where
        \begin{align}
         &  I^{A}(a,b;\xi_Z) = \int \dfrac{d^4 k}{(2\pi)^4}
         \dfrac{1}{(m_\ell^2-(k-p)^2)(\xi_Z M_Z^2-(k-a)^2)(k-b)^2}= \nonumber \\
         & \hspace{20em} = -\frac{i\pi^2}{(2\pi)^4}\!\!\int \{dx\}_3 \frac{1}{D^{A}(a,b;\xi_Z)}, \\
         &   I^{A}_{\lambda}(a,b;\xi_Z) = \int \dfrac{d^4 k}{(2\pi)^4}
         \dfrac{k_\lambda }{(m_\ell^2-(k-p)^2)(\xi_Z M_Z^2-(k-a)^2)(k-b)^2}= \nonumber \\
         & \hspace{20em} = -\frac{i\pi^2}{(2\pi)^4}\int \{dx\}_3 \frac{(xp+ya+zb)_\lambda}{D^{A}(a,b;\xi_Z)},\\
         &   I^{A}_{i\lambda}(a,b;\xi_Z) = \int \dfrac{d^4 k}{(2\pi)^4}
         \dfrac{k_i k_\lambda}{(m_\ell^2-(k-p)^2)(\xi_Z M_Z^2-(k-a)^2)(k-b)^2}= \frac{i\pi^2}{(2\pi)^4}\frac{g_{i\lambda}}{4}\ln \frac{\Lambda^2}{M_W^2} -\nonumber \\&
         - \frac{i\pi^2}{(2\pi)^4}\int \{dx\}_3 \frac{(xp+ya+zb)_i (xp+ya+zb)_\lambda}{D^{A}(a,b;\xi_Z)}
         - \frac{i\pi^2}{(2\pi)^4} \frac{g_{i\lambda}}{2}\int \{dx\}_3
         \ln\frac{D^{A}(a,b;\xi_Z)}{M_W^2}, \\
         &   D^{A}(a,b;\xi_Z) = (xp+ya+zb)^2 + y(\xi_Z M_Z^2-a^2) - zb^2.
        \end{align}
        The divergent part of this amplitude is
        \begin{equation}
            M_{fi,div}^{XZA} = - ic_\gamma\frac{e q_\ell g}{4\cos\theta_W}  \Lambda_1 \bar\ell(p^\prime) \left[ \hat{\overline{P}}_Z \gamma_\nu \gamma_\lambda \gamma_\rho -  \gamma_\rho \gamma_\lambda \gamma_\nu \hat P_Z \right] \ell(-p) \epsilon^{\lambda_X}_\mu(q)\, \epsilon^{\mu\nu\lambda\rho}.
        \end{equation}

    The amplitude of the process for the diagrams in Fig.\ref{fig:XAphiz} is given by
     \begin{multline}
             M_{fi}^{X\phi_z A} =  \frac{m_\ell q_\ell  \sin2\theta_W}{ v^2}  c_\gamma\, \bar\ell(p^\prime) \left[ \gamma^5 \left( I^{A}_\lambda (q,0;\xi_Z)  (m_\ell - \not{p}) q_\nu + I^{A}_{i\lambda} (q,0;\xi_Z)  \gamma^i q_\nu \right) \gamma_\rho + \right. \\ \left. + \gamma_\rho \left( I^{A}_\nu (0,q;\xi_Z)  (m_\ell - \not{p}) q_\lambda + I^{A}_{i\nu} (0,q;\xi_Z)  \gamma^i q_\lambda \right) \gamma^5 \right] \ell(-p) \epsilon^{\lambda_X}_\mu(q)\, \epsilon^{\mu\nu\lambda\rho},
        \end{multline}
       The divergent part of this amplitude is
        \begin{equation}
             M_{fi,div}^{X\phi_z A} =  ic_\gamma \frac{m_\ell q_\ell \sin2\theta_W }{ v^2}  \Lambda_1\, \bar\ell(p^\prime)   \gamma^5 \gamma_\lambda \gamma_\rho q_\nu  \ell(-p) \epsilon^{\lambda_X}_\mu(q)\, \epsilon^{\mu\nu\lambda\rho}.
        \end{equation}

\subsection{Vertices $XhZ$ and $XhA$}

\begin{figure}[h]
        \centering
        \includegraphics[width=0.7\textwidth]{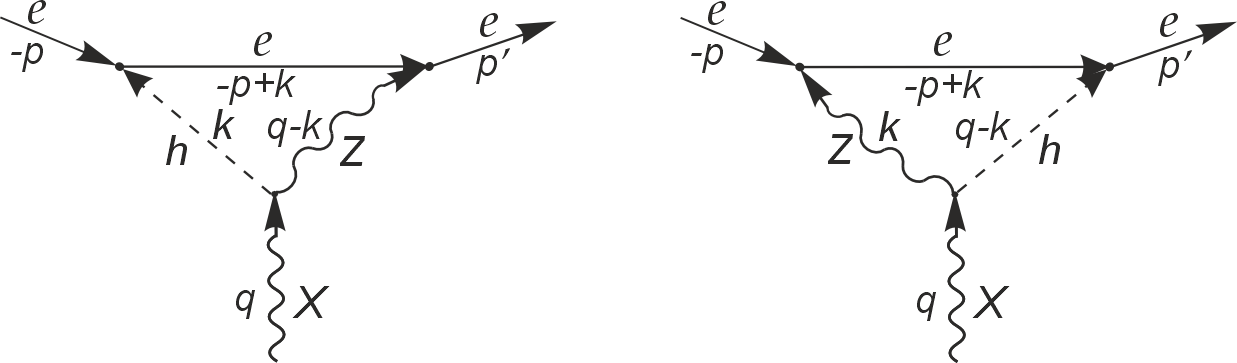} \caption{Triangle diagrams for lepton production via CS-boson decay, mediated by the interactions of the CS boson with Higgs bosons and \(Z\) bosons.}
        \label{fig:XZh}
    \end{figure}

\begin{figure}[h]
        \centering
        \includegraphics[width=0.7\textwidth]{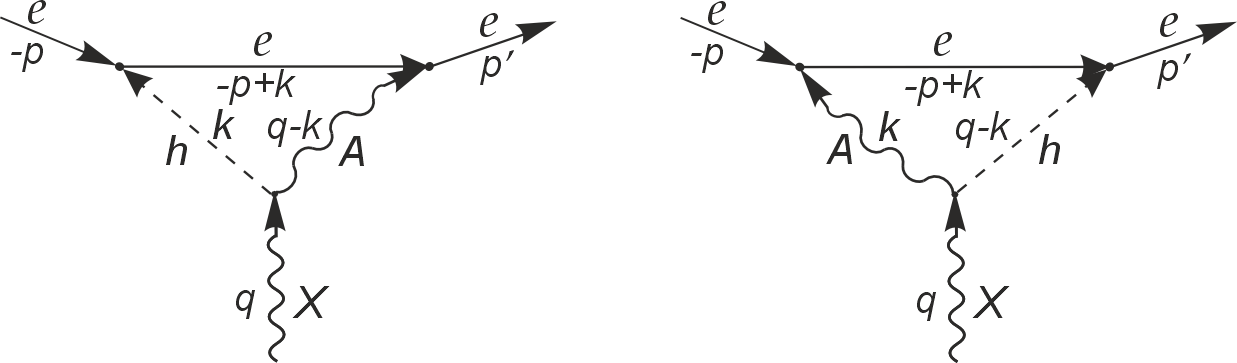} \caption{Triangle diagrams for lepton production via CS-boson decay, mediated by the interactions of the CS boson with Higgs bosons and photons}
        \label{fig:XAh}
 \end{figure}

Triangle diagrams for lepton production via CS-boson decay, mediated by the interactions of the CS boson with Higgs bosons and \(Z\) bosons, or with Higgs bosons and photons,  are presented in Fig.\ref{fig:XZh} and Fig.\ref{fig:XAh}.

The amplitude of the process for the diagrams in Fig.\ref{fig:XZh} is given by
        \begin{multline}
              M_{fi}^{XhZ} = ic_{zh} \frac{g m_\ell}{2 v^2 \cos\theta_W} \bar\ell(p^\prime) q_\lambda \gamma_\rho \hat P_Z \left[ I^{zh}_\nu (0,q)  (m_\ell - \not{p}) + \gamma^\alpha I^{zh}_{\nu\alpha} (0,q) \right] \ell(-p) \epsilon^{\mu\nu\lambda\rho} \epsilon^{\lambda_X}_\mu(q) + \\ +i c_{zh} \frac{g m_\ell}{2 v^2 \cos\theta_W} \bar\ell (p') q_\nu \left[ I^{zh}_\lambda (q,0) (m_\ell - \not{p}) + \gamma^\alpha I^{zh}_{\lambda\alpha} (q,0) \right] \gamma_\rho \hat P_Z \ell(-p) \epsilon^{\mu\nu\lambda\rho} \epsilon^{\lambda_X}_\mu(q),
        \end{multline}
        where
        \begin{align}
         &   I^{Zh}_{\lambda}(a,b) = \int \dfrac{d^4 k}{(2\pi)^4}
         \dfrac{k_\lambda }{(m_\ell^2-(k-p)^2)(m_h^2-(k-a)^2)(M_Z^2-(k-b)^2)}= \nonumber \\ 
         &\hspace{20em }=\frac{i\pi^2}{(2\pi)^4}\int \{dx\}_3 \frac{(xa+yb+zp)_\lambda}{D^{Zh}(a,b)},\\
         &   I^{Zh}_{\alpha\lambda}(a,b) = \int \dfrac{d^4 k}{(2\pi)^4}
         \dfrac{k_\alpha k_\lambda}{(m_\ell^2-(k-p)^2)(m_h^2-(k-a)^2)(M_Z^2-(k-b)^2)}= -\frac{i\pi^2}{(2\pi)^4}\frac{g_{\alpha\lambda}}{4}\ln \frac{\Lambda^2}{M_W^2} +\nonumber \\&
         + \frac{i\pi^2}{(2\pi)^4}\int \{dx\}_3 \frac{(xa+yb+zp)_\alpha (xa+yb+zp)_\lambda}{D^{Zh}(a,b)}
         + \frac{i\pi^2}{(2\pi)^4} \frac{g_{\alpha\lambda}}{2}\int \{dx\}_3
         \ln\frac{D^{Zh}(a,b)}{M_W^2}, \\
         &   D^{Zh}(a,b) = (xa+yb+zp)^2 + x(m_h^2 - a^2) + y(M_Z^2-b^2).
        \end{align}
        The divergent part of this amplitude is
        \begin{equation}
             M_{fi,div}^{XhZ} = c_{zh} \frac{ g m_\ell (t_3^\ell - 2 q_\ell \sin^2 \theta_W)}{2 v^2 \cos\theta_W}\Lambda_1 \, \bar\ell(p^\prime)  \gamma_\lambda \gamma_\rho q_\nu \, \ell(-p) \epsilon^{\lambda_X}_\mu(q)\, \epsilon^{\mu\nu\lambda\rho}.
        \end{equation}

The amplitude of the process for diagrams of Fig.\ref{fig:XAh} is given by
  \begin{multline}
              M_{fi}^{XAh} =  -ic_{\gamma h} \frac{e q_\ell m_\ell}{v^2} \bar\ell(p^\prime) q_\lambda \gamma_\rho \left[ I^{Ah}_\nu (0,q) (m_\ell - \not{p}) + \gamma^\alpha I^{Ah}_{\nu\alpha} (0,q) \right] \ell(-p) \epsilon^{\lambda_X}_\mu(q)\, \epsilon^{\mu\nu\lambda\rho} + \\ + i c_{\gamma h} \frac{e q_\ell m_\ell}{v^2} \bar\ell (p') \left[ I^{Ah}_\lambda (q,0) (m_\ell - \not{p}) + \gamma^\alpha I^{Ah}_{\lambda\alpha} (q,0) \right] \gamma_\rho q_\nu \ell(-p) \epsilon^{\lambda_X}_\mu(q)\, \epsilon^{\mu\nu\lambda\rho},
        \end{multline}
        where
        \begin{align}
         &   I^{Ah}_{\lambda}(a,b) = \int \dfrac{d^4 k}{(2\pi)^4}
         \dfrac{k_\lambda }{(m_\ell^2-(k-p)^2)(m_h^2-(k-a)^2)(k-b)^2}= -\frac{i\pi^2}{(2\pi)^4}\int \{dx\}_3 \frac{(xa+yb+zp)_\lambda}{D^{Ah}(a,b)},\\
         &   I^{Ah}_{\alpha\lambda}(a,b) = \int \dfrac{d^4 k}{(2\pi)^4}
         \dfrac{k_\alpha k_\lambda}{(m_\ell^2-(k-p)^2)(m_h^2-(k-a)^2)(k-b)^2}= \frac{i\pi^2}{(2\pi)^4}\frac{g_{\alpha\lambda}}{4}\ln \frac{\Lambda^2}{M_W^2} -\nonumber \\&
         - \frac{i\pi^2}{(2\pi)^4}\int \{dx\}_3 \frac{(xa+yb+zp)_\alpha (xa+yb+zp)_\lambda}{D^{Ah}(a,b)}
         - \frac{i\pi^2}{(2\pi)^4} \frac{g_{\alpha\lambda}}{2}\int \{dx\}_3
         \ln\frac{D^{Ah}(a,b)}{M_W^2}, \\
         &   D^{Ah}(a,b) = (xa+yb+zp)^2 + x(m_h^2 - a^2) - y b^2 .
        \end{align}
         The divergent part of this amplitude is
        \begin{equation}
             M_{fi,div}^{XAh} = c_{\gamma h} \frac{e q_\ell m_\ell}{v^2} \Lambda_1 \bar\ell(p^\prime)  \gamma_\lambda \gamma_\rho q_\nu   \ell(-p) \epsilon^{\lambda_X}_\mu(q)\, \epsilon^{\mu\nu\lambda\rho}.
        \end{equation}

\section{Divergences in the diagrams with a circular loop}\label{sec:circular}

\begin{figure}[h!]\centering
            \includegraphics[width = 0.65\textwidth]{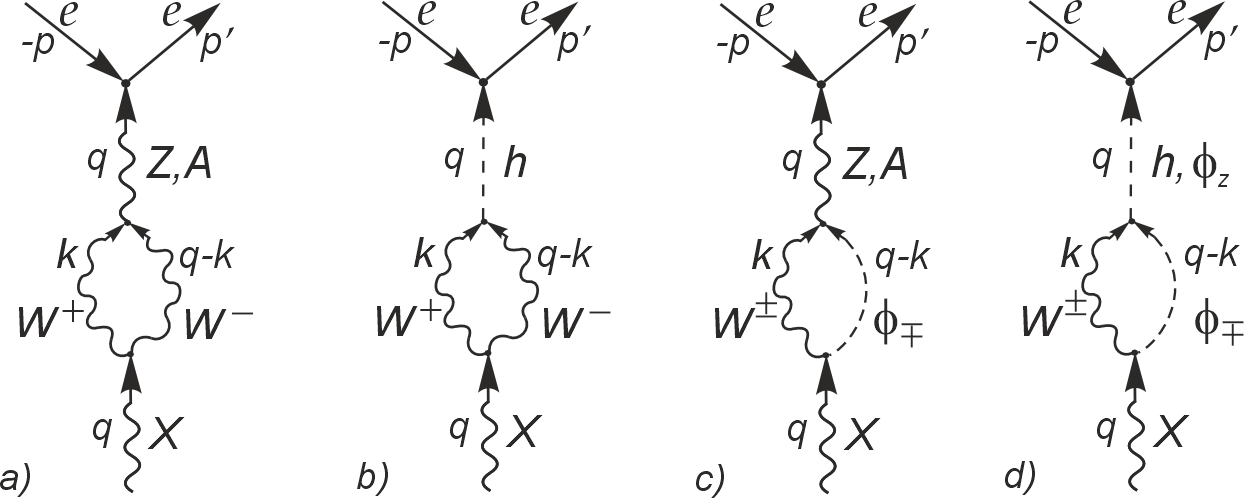}
            \captionsetup{width=0.9\textwidth}
            \caption{Loop diagrams for lepton production via CS-boson decay, mediated by the interactions of the CS boson with two $W$ bosons or with a $W$ and a charged $\phi$ boson.}
            \label{fig:Rxi_circular_diag}
        \end{figure}

Loop diagrams for lepton production via CS-boson decay, mediated by the interactions of the CS boson with two $W$ bosons or with a $W$ and a charged $\phi$ boson, are shown in Fig.\ref{fig:Rxi_circular_diag}.

Let us consider, as an example, diagram~\textit{a}) in Fig.~\ref{fig:Rxi_circular_diag}, mediated by a virtual \(Z\) boson
              \begin{multline}\label{XWWZ}
           M_{fi}^{XWWZ}  =   - \dfrac{g^4}{2} \frac{\bar{\ell}(p^\prime) \gamma_\rho (g_V^f - g_A^f \gamma_5) \ell(-p)}{(q^2-m_Z^2)} q_\lambda   \varepsilon_\mu^{\lambda_X}(q) \varepsilon^{\mu \nu \lambda \rho} \int \dfrac{d^4 k}{(2\pi)^4}  
                \dfrac{k_\nu }{((k-q)^2-m_W^2)(k^2-m_W^2)}  \\
                 \left(c_2 \left(3-(1-\xi_W)\dfrac{(q-k)\cdot (q+k)}{(q-k)^2-\xi_W m_W^2}\right) +
            c_2^* \left(3-(1-\xi_W)\dfrac{k\cdot (2q-k)}{k^2-\xi_W m_W^2}\right)\right) =\\=
            -3(c_2 +c_2^* ) \dfrac{g^4}{2} \frac{\bar{\ell}(p^\prime) \gamma_\rho (g_V^f - g_A^f \gamma_5) \ell(-p)}{(q^2-m_Z^2)} q_\lambda   \varepsilon_\mu^{\lambda_X} (q) \varepsilon^{\mu \nu \lambda \rho}
                  Y_{a,\nu}(1)+\\+ (1-\xi_W)\left[A_\rho\, Y_{b,\nu}(\xi_W)+ B_\rho\, Y_{c,\nu}(\xi_W)\right]q_\lambda   \varepsilon_\mu^{\lambda_X} (q) \varepsilon^{\mu \nu \lambda \rho},
            \end{multline}
where   $Y_{a,\nu}(\xi_W)$ contains divergence  and is proportional to $q_\nu$ 
\begin{multline}\label{IntYa}
            Y_{a,\nu}(\xi_W)=\int\dfrac{d^4 k}{(2\pi)^4}   \dfrac{k_\nu }{((q-k)^2-m_W^2)(k^2 - \xi_W m_W^2)}=\\=
            \frac{i \pi^2}{(2\pi)^4}q^\nu\left[ \frac{1}{2} \ln \frac{\Lambda^2}{M_W^2}-\int\limits_0^\infty dx\,x 
            \ln\frac{x(1-x)\frac{q^2}{M_W^2}+x(1-\xi)+\xi}{x(1-\xi)+\xi}\right],
        \end{multline}
and $Y_{b,\nu}(\xi_W)$, $Y_{c,\nu}(\xi_W)$ do not contain divergences 
and are also proportional to $q_\nu$ 
        \begin{equation}\label{IntYb}
            Y_{b,\nu}(\xi_W)=\int \dfrac{d^4 k}{(2\pi)^4} 
                \dfrac{ k_\nu}{((k-q)^2-m_W^2)(k^2-m_W^2)}\cdot \dfrac{(q-k)\cdot (q+k)}{(q-k)^2-\xi_W m_W^2}\sim q_\nu,
        \end{equation}
        \begin{equation}\label{IntYc}
            Y_{c,\nu}(\xi_W)=\int \dfrac{d^4 k}{(2\pi)^4} 
                \dfrac{ k_\nu}{((k-q)^2-m_W^2)(k^2-m_W^2)}\cdot \dfrac{k\cdot (2q-k)}{k^2-\xi_W m_W^2} \sim q_\nu.
        \end{equation}
Due to the presence of the Levi-Civita tensor in Eq.~\eqref{XWWZ}, 
the contributions $Y_{a,\nu}(\xi_W)$, $Y_{b,\nu}(\xi_W)$, and 
$Y_{c,\nu}(\xi_W)$ vanish. Consequently, the amplitude of diagram~\textit{a}) 
in Fig.~\ref{fig:Rxi_circular_diag} is zero, $M_{fi}^{XWWZ} = 0$.
Similarly, one can show that the amplitudes of all diagrams in 
Fig.~\ref{fig:Rxi_circular_diag} vanish in the $R_\xi$ gauge.

\begin{figure}[b]
        \centering
        \includegraphics[width=0.29\textwidth]{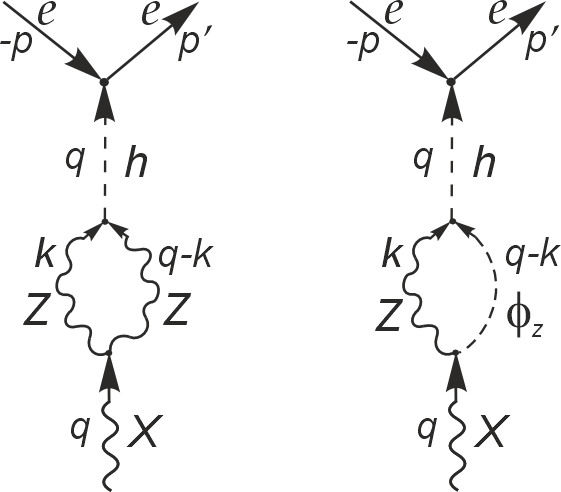}
        \caption{Loop diagrams for lepton production via CS-boson decay, mediated by the interactions of the CS boson with two $Z$ bosons or with a $Z$ and a neutral $\phi_Z$ boson.}
        \label{fig:XZZ_circle_Rxi}
        \end{figure}

Loop diagrams for lepton production via CS-boson decay, mediated by the interactions of the CS boson with two $Z$ bosons or with a $Z$ and a neutral $\phi_Z$ boson, are shown in Fig.\ref{fig:XZZ_circle_Rxi}. It can be shown that the amplitudes of these loop diagrams also vanish in the \(R_\xi\) gauge due to the antisymmetric properties of the Levi–Civita symbol.

If the Higgs field were charged under the group $U_X(1)$, additional diagrams would arise due to the mixing between the Stuckelberg field and the Higgs Goldstone bosons induced by the gauge fixing of the $X_\mu$ field, see, e.g.,~\cite{Ruegg:2003ps}. However, according to Table~4 of~\cite{Antoniadis:2009ze}, the Higgs field is neutral under $U_X(1)$. 

Finally, diagrams involving ghost fields cannot be constructed, since none of the SM ghost fields couple to the Stuckelberg field $X_\mu$.

Thus, we have evaluated all triangle and circular loop diagrams presented in Sections~\ref{sec:triadagr} and~\ref{sec:circular} that describe the interaction of the CS boson with leptons, and we can conclude that no additional diagrams give non-zero contributions.

\section{Lack of cancellation of divergent terms in the loop \\ diagrams in $R_\xi$ gauge}

The sum of the divergent parts of all the above-considered diagrams can be presented as
      \begin{equation}\label{divpartL}
            \sum_{diagrams} M_{fi,div} =\Lambda_1  \bar \ell(p^{\prime}) \left[(A+A_5 \gamma^5)\, \gamma_\nu \gamma_\lambda \gamma_\rho  + \frac{B+B_5 \gamma^5}{v}\, q_\nu\gamma_\lambda \gamma_\rho  \right] \ell(-p)\varepsilon_\mu^{\lambda_X}(q)\,\epsilon^{\mu\nu\lambda\rho},
        \end{equation}
where $A$, $A_5$, $B$, $B_5$ are dimensionless factors, namely
\begin{multline}
\label{coefA1}
    A+A_5 \gamma^5 =\\  -i\frac{g^2}{4} \left[\Theta_{W1} + \frac{2 c_z}{\cos^2\theta_W}   \left(  t_3^\ell \left( t_3^\ell- 2 q_\ell \sin^2\theta_W \right) + 2 q_\ell^2 \sin^4\theta_W \right) + 2q_\ell  c_{\gamma}\tan\theta_W   \left( t_3^\ell-2q_\ell \sin^2\theta_W \right) \right]\\
    - i\gamma^5 \frac{g^2}{4} \left[\Theta_{W1} + \frac{2t_3^\ell  c_z}{\cos^2\theta_W}    \left( t_3^\ell - 2 q_\ell \sin^2\theta_W \right) + 2q_\ell t_3^\ell  c_{\gamma}\tan\theta_W   \right],
\end{multline}
        \begin{multline}\label{coefB1}
            B+B_5 \gamma^5 =-  \frac{m_\ell}{v} \left[   \Theta_{W2}   - c_{zh} \frac{ g (t_3^\ell - 2 q_\ell \sin^2 \theta_W)}{2 \cos\theta_W} -  c_{\gamma h}  e q_\ell \right]-\\-
           i \frac{m_\ell}{v} \gamma^5\left[    \Theta_{W1}  - c_z (t_3^\ell-2q_\ell \sin^2\theta_W)   - q_e c_\gamma \sin2\theta_W   \right]. 
        \end{multline}
The condition for the absence of divergences is obviously $A=A_5=B=B_5=0$.

Conditions $A=A_5=0$ gives
    \begin{align}
            &c_z =  \cot \theta_W \, c_\gamma, \label{Aeq1} \\
            &\Theta_{W1} = -\frac{2}{g} \frac{t_3^\ell}{\cos\theta_W} \left( \frac{g}{2} \frac{2(t_3^\ell-2q_\ell \sin^2\theta_W)}{\sin\theta_W} + q_\ell e \right) c_\gamma.\label{Aeq2}
    \end{align}
Conditions $B=B_5=0$, with taking into account the relation \eqref{Aeq1}, gives
        \begin{align}
        &\Theta_{W1} =  c_z (t_3^\ell-2q_\ell \sin^2\theta_W)   + q_e c_\gamma \sin2\theta_W, \label{Deq2}\\
     &\Theta_{W2} = c_{zh} \frac{g (t_3^\ell - 2 q_\ell \sin^2 \theta_W)}{2 \cos\theta_W} + c_{\gamma h} \, (e q_\ell). \label{Deq1} 
        \end{align}
Unfortunately, there are no non-trivial values of the theory's coupling parameters ($\Theta_{W1}$, $\Theta_{W2}$, $c_z$, $c_\gamma$, $c_{\gamma h}$ and $c_{z h}$) that would satisfy the condition of absence of divergences  $A=A_5=B=B_5=0$.

As one can see, only one relation, \eqref{Deq1}, involves the couplings 
$\Theta_{W2}$, $c_{zh}$, and $c_{\gamma h}$. 
This implies that if the dimension-five operators describing the interaction of the CS boson with the Higgs field in the initial effective Lagrangian \eqref{Lcs5} are neglected, i.e., if we set 
$c_{zh}=c_{\gamma h}=0$, and consider only the dimension-4 interactions of the CS boson with the SM vector fields in Lagrangian \eqref{Lcs}, then Eq.\eqref{Deq1} leads to the condition $\Theta_{W2}=0$. Therefore, the parameter $\Theta_{W2}$ plays a distinctive role in the theory. 

If we rewrite the interaction Lagrangian~\eqref{Lcs} in the absence of the Higgs field in the form
\begin{equation}\label{Lcs1}  
     \mathcal{L}_{CS}=c_z X_\mu J_{ZZ}^\mu  +c_\gamma X_\mu J^\mu_{ZA} +\Theta_{W1}  X_\mu  J^{-,\mu}_{WW}+i \Theta_{W2}  X_\mu  J^{+,\mu}_{WW},
\end{equation}
where
\begin{align}
   &  J_{ZZ}^\mu = \epsilon^{\mu\nu\lambda\rho} Z_\nu \partial_\lambda Z_\rho, \quad J^\mu_{ZA} = \epsilon^{\mu\nu\lambda\rho}  Z_\nu \partial_\lambda A_\rho, \label{curZZ}\\
   &  J^{-,\mu}_{WW} =\epsilon^{\mu\nu\lambda\rho} (W_\nu^- \partial_\lambda W_\rho^+ - W_\rho^+ \partial_\lambda W_\nu^-)=\tilde W^{\mu\nu+}W_\nu^- +  \tilde W^{\mu\rho-}W_\rho^+,\label{curWWm}\\
   &  J^{+,\mu}_{WW} = \epsilon^{\mu\nu\lambda\rho}( W_\nu^- \partial_\lambda W_\rho^+ + W_\rho^+ \partial_\lambda W_\nu^-)=\tilde W^{\mu\nu+}W_\nu^- -  \tilde W^{\mu\rho-}W_\rho^+\label{curWWp},\\
   & W_{\lambda\rho}^\pm = \partial_\lambda W_\rho^\pm - \partial_\rho W_\lambda^\pm, \quad \tilde W^{\mu\nu\pm}= \frac12\epsilon^{\mu\nu\lambda\rho} W_{\lambda\rho}^\pm.
\end{align}
Then 
\begin{align}
   &  \partial_\mu J_{ZZ}^\mu = \frac12 \tilde Z^{\mu\nu} Z_{\mu\nu}, \quad \partial_\mu J^\mu_{ZA} = \frac12 \tilde Z^{\mu\nu} A_{\mu\nu} = \frac12 \tilde A^{\mu\nu} Z_{\mu\nu},  \label{DcurZZ}\\
   & \partial_\mu J^{-,\mu}_{WW} = \tilde W^{\mu\nu-
   }W_{\mu\nu}^+ =  W^{\mu\nu-} \tilde W_{\mu\nu}^+, \quad \partial_\mu J^{+,\mu}_{WW} = 0,\label{DcurWWm}
\end{align}
where $Z_{\mu\nu}=\partial_\mu Z_\nu-\partial_\nu Z_\mu$, $A_{\mu\nu}= \partial_\mu A_\nu-\partial_\nu A_\mu$, $\tilde Z^{\mu\nu}= \frac12\epsilon^{\mu\nu\lambda\rho} Z_{\lambda\rho}^\pm$, and $\tilde A^{\mu\nu}= \frac12\epsilon^{\mu\nu\lambda\rho} A_{\lambda\rho}^\pm$. As one can see,  the parameter $\Theta_{W2}$ is associated with the current $J^{+,\mu}_{WW}$, which is the only conserved current in the Lagrangian \eqref{Lcs1}.

Let us now consider the problem of canceling divergent terms within an alternative approach. 
In this paper, we restrict ourselves to considering the effective interaction of the CS boson with SM particles in the form of the dimension-six operators \eqref{L1} and \eqref{L2}. In this case, the conditions $A=0$, $A_5=0$, $B=0$, and $B_5=0$ should be rewritten using \eqref{cgammah}--\eqref{L9}, and they then involve only four independent parameters of the theory ($c_2$, $\tilde c_2$, $c_1$, $\tilde c_1$). We then obtain the following conditions
  \begin{align}
            & \tilde c_1 = c_2 \frac{g}{2} \, \frac{g-g'\cot\theta_W}{g\cot\theta_W + g'} = 0,\label{cond1} \\
            & \tilde c_1 =   \frac{c_2}{2} \, \left( -\frac{g}{2 t_3^\ell} \frac{\sin2\theta_W}{t_3^\ell - q_\ell s^2_W} - g' \right),\\
            & \tilde c_1 = c_2 \frac{g}{2} \, \frac{g - g' q_\ell \sin 2\theta_W - g(t_3^\ell - 2 q_\ell \sin^2\theta_W)}{g q_\ell \sin 2\Theta_W - g'(t_3^\ell - 2 q_\ell \sin^2\theta_W)}, \label{Deq0form2} \\
            & c_1 = \tilde c_2  \frac{g}{2} \, \frac{1 + t_3^\ell}{q_\ell \sin 2\theta_W - \tan\theta_W (t_3^\ell - 2 q_\ell \sin^2\theta_W)}.\label{cond4}
\end{align}
Thus, we obtain the first three relations between $c_2$ and $\tilde c_1$, which are satisfied only in the trivial case $c_2 = \tilde c_1 = 0$. 
Moreover, generalizing the result for the interaction with all SM fermions of the same flavors, the last relation cannot be satisfied simultaneously for all SM fermions (charged leptons, neutrinos, top and bottom quarks) since it depends both on the third component of the isospin $(t_3^f)$ and on the electric charge $(q_f)$ of the fermion.
 Therefore, in this approach, there is also no non-trivial choice of the theory’s coupling parameters that would satisfy the condition for the cancellation of divergences, i.e., $ A = A_5= B = B_5=0$.

The expressions considered in our paper involve linearly divergent integrals, which, similarly to the case of the chiral anomaly, see, e.g., \cite{Cheng:1984vwu}, are sensitive to shifts of the integration variable by a constant. However, such a shift changes the value of the integrals only by a finite amount. As a result, the expressions for $A$, $A_5$, $B$, and $B_5$ remain unaffected.

It should be noted that conditions \eqref{Aeq1} -- \eqref{Deq1} or \eqref{cond1} -- \eqref{cond4} do not depend on any gauge parameters $\xi_i$. This rules out the possibility that there exist values of the gauge parameters $ (\xi_i) $ for which the divergences cancel, as happens with infrared divergences in QED in the Fried–Yennie gauge \cite{Fried:1958zz}.

Thus, the divergent part of the effective interaction of the CS bosons with leptons of SM \eqref{divpartL}  derives the following terms in the Lagrangian, involving dimension-four and dimension-five operators
        \begin{equation}\label{Ldivpart}
            \mathcal{ L}^{int,div}_{X\ell \ell}= \bar\ell  \gamma^\mu (\alpha_\ell^{div} +\beta_\ell^{div} \gamma^5) \ell X_\mu +\frac{m_\ell}{v^2}  \bar\ell  \sigma^{\mu\nu}( \gamma_\ell^{div}+\delta_\ell^{div} \gamma^5)  \ell  {X}_{\mu\nu},
        \end{equation}
where $\alpha_\ell^{div}$,  $\beta_\ell^{div}$, $\gamma_\ell^{div}$, $\delta_\ell^{div}$ are dimensionless parameters containing divergences, $X_{\mu\nu}= \partial_\mu X_\nu-\partial_\nu X_\mu$, and we have used relations
\begin{equation}\label{e3gamma}
\epsilon^{\alpha\beta\gamma\delta}\sigma_{\alpha\beta}=-2i\gamma^5 \sigma^{\gamma\delta},
 \quad \epsilon^{\alpha\mu\nu\rho}\gamma_\mu \gamma_\nu \gamma_\rho= 6 i\gamma^\alpha \gamma^5.
 \end{equation}

 The resulting divergent terms \eqref{divpartL} are in full agreement with our previous calculations performed in the unitary gauge \cite{Borysenkova:2024xno}, see Appendix \ref{AppD}.
 Thus, we conclude that the Lagrangian \eqref{Lcs}, which contains the interaction operators of vector fields of dimension-4 and seems to be renormalizable, is unfortunately non-renormalizable.

This result is not surprising, since the initial interaction of the CS bosons with SM fields is given by dimension-six operators \eqref{L1} and \eqref{L2}. We conclude that after the electroweak symmetry breaking, the interaction of the CS boson with fermions of the same flavor should be considered within the framework of the effective field theory approach \cite{Buchmuller:1985jz,Georgi:1991ch,Georgi:1992xg,Brivio:2017vri} and can be written as
\begin{equation}\label{Leffff}
            \mathcal{ L}^{eff}_{Xff}= \bar f  \gamma^\mu (\alpha_f +\beta_f \gamma^5) f X_\mu +\frac{m_f}{v^2}  \bar f  \sigma^{\mu\nu}(\gamma_f+\delta_f \gamma^5)  f  {X}_{\mu\nu}+ \mathcal{L}'_{X ff},
\end{equation}
where dimensionless parameters $\alpha_f$, $\beta_f$, $\gamma_f$, $\delta_f$ should be considered as new couplings of the effective theory, and $\mathcal{L}'_{Xff}$ is a well-defined Lagrangian of interaction depending on the coupling parameters of the  Lagrangians \eqref{Lcs}, \eqref{Lcs5}. It can be seen that this effective interaction becomes multi-parametric, making it analogous in this respect to the effective interaction of axion-like particles with SM particles, see, e.g., \cite{Alekhin:2015byh}.

The next section is devoted to constructing the effective interaction Lagrangian and discussing the number of independent parameters that characterize it.

\section{Effective interaction between vector Chern-Simons bosons and fermions}

In the previous section, it was shown that the divergent part of the loop diagrams describing the interaction between fermions and Chern–Simons bosons do not cancel and that their explicit expressions, including all coefficients, are independent of the gauge choice.  In this section, we are interested in the expressions for the full loop interaction amplitude, including the finite part. This will help us describe the loop-induced interaction of the GeV-scale CS boson with SM fermions within the framework of the effective field theory formalism and determine the number of independent parameters of the effective interaction.

To simplify the calculation, we perform it in the unitary gauge, considering, for definiteness, the decay of the GeV-scale Chern–Simons boson into two fermions of the same flavor. For a complete and rigorous treatment, we should have taken into account all amplitudes corresponding to the following interactions at the vertex with the CS boson:
\begin{equation}
\label{AmplGeneral}
M_{fi}^{X \rightarrow f \bar f} = M_{fi}^{XWW} + M_{fi}^{XZZ}+M_{fi}^{XZh} + M_{fi}^{XZA} +M_{fi}^{XAh}.
\end{equation}
However, for practical reasons, we restrict ourselves to leading order, including only those terms in both convergent and divergent parts that are not suppressed by the electroweak scale mass, such as terms proportional to $\sim 1/M_W$. Therefore, the contributions from diagrams involving the $XZh$ and $XAh$ vertices are neglected. Such an approximation cannot be used to describe the interaction of the CS boson with the top quark. The interaction with the top quark requires separate, detailed consideration.

In Appendices \ref{AppE}, \ref{AppF}, and \ref{AppG}, we consider in detail the decay of the GeV-scale CS boson into a pair of leptons and into a pair of quarks of the same flavor. 
As a result, we obtain the expression for the full amplitude of the decay of the CS boson into any fermion-antifermion pair via all possible channels at leading order. As it turns out, they look the same as the first term of relation \eqref{Leffff}. 
Corrections in the form of higher-dimensional operators enter either into the coefficients $\gamma_f$ and $\delta_f$, or into the Lagrangian $\mathcal{L}'_{X ff}$ in \eqref{Leffff}. We have also shown that the Lagrangian $\mathcal{L}'_{X ff}$ is composed entirely of higher-dimensional operators. Consequently, both these corrections and the Lagrangian $\mathcal{L}'_{X ff}$ itself will not be considered in the following analysis.

Therefore, at leading order, the amplitude for the decay of the CS boson into a pair of leptons or a pair of quarks of the same flavor takes the following form
\begin{equation}
\label{fullamplitude}
\begin{split}
    M_{fi}^{X \rightarrow f \bar f} &= \bar f (p') \left[ \frac{g^2}{2} \Theta_{W1} A_f^W \gamma^\mu \gamma^5 \hat P_L \right] f (-p) \, \epsilon^{\lambda_X}_\mu (q) + \\ &\phantom{aa} + \bar f (p') \left[ \frac{g^2}{4 \cos^2\theta_W} c_z A_f^Z \gamma^\mu \gamma^5 \hat P_Z^2 \right] f (-p) \, \epsilon^{\lambda_X}_\mu (q) + \\ &\phantom{aa} + \bar f (p') \left[ \frac{g e q_f}{2 \cos\theta_W} c_\gamma A_f^\gamma \gamma^\mu \gamma^5 \hat P_Z \right] f (-p) \, \epsilon^{\lambda_X}_\mu (q),
\end{split}
\end{equation}
where the coefficients $A_f^V$ determine the contribution from loop diagrams with the corresponding vector bosons $V$, and the operator $\hat P_Z$ was defined in \eqref{PSdec}. This result was derived explicitly in Appendices~\ref{AppE}, \ref{AppF}, and \ref{AppG}.

Assuming a GeV-scale mass for the CS boson, $M_W^2\gg M_X^2$, and taking into account that the $M_W$-boson mass is much larger than the lepton and quark masses (except for the top quark), $M_W^2\gg m_\ell^2$, $m_q^2$, we obtain
\begin{equation}\label{aW1}
    A_f^W= 6\hat{\Lambda}_{1f}^{W} =
-\frac{6 \pi^2}{(2\pi)^4} \int_0^{1}dx (1-x) \ln\frac{\Lambda^2 x}{M_W^2(1-x)},\quad f=\ell,\, \nu,\, u,\,c
\end{equation}
\begin{multline}\label{aW2}
    A_f^W= 6 \sum_{i=u,c,t} V_{fi}^\dagger V_{if} \, \hat \Lambda_{1i}^W =
-\frac{6\pi^2}{(2\pi)^4} \int_0^{1}dx (1-x)\times \\ \left\{(\vert V_{uf} \vert^2 + \vert V_{cf} \vert^2) \ln\left[\frac{\Lambda^2}{M_W^2}\frac{x}{1-x}\right]  + \vert V_{tf} \vert^2 \ln\left[\frac{\Lambda^2}{M_W^2}\frac{x}{1+\left(\frac{m_t^2}{M_W^2}-1\right)x} \right]  \right\},\quad f=d,\,s,\,b
\end{multline}
\begin{align}
  &  A_f^Z= 6\hat{\Lambda}_{1f}^{Z} =
-\frac{6 \pi^2}{(2\pi)^4} \int_0^{1}dx (1-x) \ln\frac{\Lambda^2 x}{M_Z^2(1-x)}, \quad f=\ell,\, \nu,\, u,\,c,\,d,\,s,\,b \label{aZ}\\
 &   A_f^\gamma = 6 (\hat \Lambda_{1f}^{(1\gamma)} + \hat \Lambda_{1f}^{(2\gamma)}) = -\frac{6\pi^2}{(2\pi)^4} \int_0^{1}dx \, (1-x) \, \left\{ 2 + 2\ln\left[\frac{\Lambda^2}{M_Z^2}\frac{x}{1-x}\right] \right\} ,\quad f=\ell,\, \nu,\, u,\,c,\,d,\,s,\,b \label{agamma}
\end{align}

The expressions \eqref{aW1}--\eqref{agamma} for the interaction coefficients contain a cut-off parameter $\Lambda$. Within the framework of effective field theory, these formally infinite coefficients are replaced by finite effective couplings. One may reasonably assume that if the corresponding coefficients are described by identical mathematical expressions, they will likely remain equal within the effective field theory approach as well.  We assume that, in leading order, for describing interactions of the CS bosons with fermions of the same flavors (all leptons and quarks except for the top quark), Lagrangian \eqref{Leffff} can be simplified to the form 
\begin{equation}\label{Leffffsimpli}
            \mathcal{ L}_{Xff}= \sum_{f}\bar f  \gamma^\mu (\alpha_f +\beta_f \gamma^5) f X_\mu,
\end{equation}
where, after expanding \eqref{fullamplitude}, we define $\alpha_f , \, \beta_f$ as
\begin{align}
    & \alpha_f = -\frac{g^2}{4} \Theta_{W1} A_f^W - \frac{g^2}{2 \cos^2\theta_W} c_z A_f^Z \, t_3^f (t_3^f - 2q_f \sin^2 \theta_W) - \frac{g e}{2 \cos\theta_W} c_\gamma A_f^\gamma \, q_f t_3^f,\label{alphaf} \\
    & \beta_f = \frac{g^2}{4} \Theta_{W1} A_f^W + \frac{g^2}{2 \cos^2\theta_W} c_z A_f^Z \left[ t_3^f (t_3^f - 2q_f \sin^2 \theta_W) + 2q_f^2 \sin^4 \theta_W \right] + \nonumber \\ &\hspace{2em} + \frac{g e}{2 \cos\theta_W} c_\gamma A_f^\gamma \, q_f (t_3^f - 2q_f \sin^2 \theta_W).\label{betaf}
\end{align}
  {It can be seen that for neutrinos, for which $q_f=0$, one automatically obtains $\alpha_f=-\beta_f$, as expected, since neutrinos are left-handed particles.}

After analyzing the expressions for coefficients \eqref{aW1}, \eqref{aZ}, \eqref{agamma}  for leptons and up-type quarks, we can assume that for particles with identical charges and isotopic spins (for which the operator $\hat P_Z$ has the same form), the couplings $\alpha_f$ and $\beta_f$ in \eqref{Leffff} are expected to be approximately equal up to a factor \((1+\mathcal{O}(m_f^2/M_W^2))\). Therefore, Lagrangian \eqref{Leffffsimpli} exhibits an approximate universality of couplings within each fermion family: the same parameters $\alpha_\nu$ apply to all three types of neutrinos $(\nu_\ell)$; $\alpha_L$ and $\beta_L$ are approximately the same for all three charged leptons $(\ell_n)$; and $\alpha_U$ and $\beta_U$ are common, within this approximation, to both $u$ and $c$ quarks.

 For down-type quarks, the couplings $\alpha_D$ and $\beta_D$ can be taken to be approximately equal in leading order for the $d$ and $s$ quarks, since the corresponding CKM combinations 
$\lvert V_{uf}\rvert^2+\lvert V_{cf}\rvert^2$ and $\lvert V_{tf}\rvert^2$ in \eqref{aW2} are nearly identical for $f=d,s$. In contrast, the $b$ quark generally requires its own independent couplings $\alpha_b$ and $\beta_b$.

Therefore, our analysis indicates that, rather than having 22 independent couplings (two for each of the six leptons and five quarks), the effective Lagrangian \eqref{Leffffsimpli} in the leading approximation actually involves only 9 independent couplings   {and can be presented in the form
\begin{multline}\label{Leffff10}
            \mathcal{ L}_{Xff}= \sum_{\ell=e,\mu,\tau}\bar \ell  \gamma^\mu (\alpha_L +\beta_L \gamma^5) \ell\, X_\mu+\alpha_{\nu}\sum_{\ell=e,\mu,\tau}\bar \nu_\ell  \gamma^\mu P_L \nu_\ell\, X_\mu+\\
            +\!\! \sum_{u_n=u,c}\bar u_n  \gamma^\mu (\alpha_U +\beta_U \gamma^5) u_n\, X_\mu+ \!\!\sum_{d_n=d,s}\bar d_n  \gamma^\mu (\alpha_D +\beta_D \gamma^5) d_n X_\mu+ \bar b  \gamma^\mu (\alpha_b +\beta_b \gamma^5) b\, X_\mu.
\end{multline}
}

These new couplings can be expressed as specific combinations of the original parameters $\Theta_{W1}$, $c_z$, and $c_\gamma$ appearing in the initial Lagrangians \eqref{L1} and \eqref{L2}.  
For example, by isolating the divergent term $\ln\frac{\Lambda^2}{M_W^2}$ within the logarithmic expressions under the integral in \eqref{aW1} -- \eqref{agamma} and discarding it using the minimal subtraction scheme \cite{tHooft:1972tcz,Peskin:1995ev} (see also Appendices~\ref{AppE} -- \ref{AppG}), we obtain finite values for the parameters $A^W_i$, $A^Z_i$, and $A^\gamma_i$. Using relation \eqref{fullamplitude}, the corresponding effective couplings for the Lagrangian \eqref{Leffffsimpli} in the leading approximation are then determined. In particular, for charged leptons and neutrinos, we obtain
\begin{align}
    & \alpha_L = -g^2 \left[ 4.75 \cdot 10^{-3} \, \Theta_{W1} + 4.47 \cdot 10^{-4} \, c_z + 1.28 \cdot 10^{-3} \, c_\gamma \right], \\
    & \beta_L = g^2 \left[ 4.75 \cdot 10^{-3} \, \Theta_{W1} + 1.95 \cdot 10^{-3} \, c_z + 1.49 \cdot 10^{-4} \, c_\gamma \right], \\
    & \alpha_\nu = -g^2 \left[ 4.75 \cdot 10^{-3} \, \Theta_{W1} + 3.84 \cdot 10^{-3} \, c_z \right].
\end{align}

  {
By analyzing the explicit expressions \eqref{alphaf} and \eqref{betaf} for the coefficients of the effective Lagrangian \eqref{Leffff10} obtained in the one loop approximation, one can identify an interesting feature. In the case where the imaginary part of the coefficient $C_Y$
 associated with the dimension-six operator $\mathcal{L}_1$ \eqref{L1} vanishes, $\tilde c_1=0$, the differences between the effective vector couplings become ultraviolet finite, at least in the leading approximation. Using the relations
 \begin{equation}
     \tan\theta_W c_z = c_\gamma,
\quad
\Theta_{W1} = c_z,
 \end{equation}
 implied by Eqs.\eqref{L7}--\eqref{L9} for $\tilde c_1=0$, we obtain
\begin{equation}\label{alphadif}
   \alpha_U - \alpha_D\approx \frac{\alpha_L - \alpha_\nu}3 =  \frac{\tan^2 \theta_W}{32 \pi^2}  \,g^2 \Theta_{W1}
\end{equation} 
The difference $\alpha_U-\alpha_D$ is of particular interest because it determines the coefficient of the neutral isovector vector current of the first-generation quarks,
\begin{equation}
J^{quark}_\mu = \alpha_u\, \bar u \gamma_\mu u + \alpha_d\, \bar d \gamma_\mu d 
= \frac{\alpha_u + \alpha_d}{\sqrt{2}}\, j_\mu^{V,s}
+ \frac{\alpha_u - \alpha_d}{\sqrt{2}}\, j_\mu^{V,0},
\end{equation}
where
\begin{equation}
j_{\mu}^{V,s} = \frac{1}{\sqrt{2}}
(\bar{u}\gamma_\mu u + \bar{d}\gamma_\mu d),
\quad
j_{\mu}^{V,0} = \frac{1}{\sqrt{2}}
(\bar{u}\gamma_\mu u - \bar{d}\gamma_\mu d)
\end{equation}
denote the isoscalar and isovector vector quark currents, respectively. Consequently, the finite result obtained above determines the isovector component of the effective CS-boson coupling to light quarks through the coupling $\Theta_{W1}$. This result may be useful for future studies of hadronic decay modes of the CS boson. We note that the same coupling $\Theta_{W1}$ also governs the production of the CS boson in heavy-meson decays \cite{Borysenkova:2021ydf}.
}

For a complete description of the loop-induced interactions of the CS bosons with SM fermions, it is also appropriate to present here the Lagrangian describing the interactions of the CS bosons with flavor-changing neutral currents at leading order, i.e., with quarks of different flavors \cite{Borysenkova:2021ydf,Gorkavenko:2024ivy}:
\begin{equation}\label{Mfi5m}
    \mathcal{L}_{Xqq'}=  \sum_{m< n}\Theta_{W1}\left(\! C_{mn}\, \overline{d_m}\, \gamma^{\mu}\,\hat
 P_L  \,  d_n X_{\mu}+C_{nm}^+ \overline{d_n}\, \gamma^{\mu}\,\hat
 P_L  \,  d_m X_{\mu}\!\right),
\end{equation}
where $d_n$ is a down-type quark of the $n$ generation, the summation occurs over the quark generations,
\begin{equation}\label{Cdm}
    C_{mn}= \frac{3a}{2\sqrt{2}\pi^2}\, G_F m_t^2\,V_{t d_m}^* V_{t d_n}
\end{equation}
and 
$
 a=0.13$,  $|C_{sb}| = 1.97\cdot 10^{-4}$,
 $|C_{db}|=   4.43\cdot 10^{-5}$,    $|C_{ds}|=   1.77\cdot 10^{-6}$. We note that this interaction is determined by only a single unknown parameter, $\Theta_{W1}$. It has been shown that the interaction of a GeV-scale CS boson with up-type quarks is negligible.

 The fact that the Lagrangian \eqref{Mfi5m} contains the factor \((1-\gamma^5)\), whereas the Lagrangian \eqref{Leffff10} involves the factor $(\alpha_f+\beta_f \gamma^5)$, is explained by the fact that the interaction in Lagrangian \eqref{Mfi5m} is generated solely by triangle diagrams with two virtual \(W\)-bosons, while the interaction in Lagrangian \eqref{Leffff10} arises from triangle diagrams involving all SM vector bosons as virtual particles.

We also draw your attention to the fact that possible ambiguities in the values of the coefficients of the effective Lagrangian, which may arise in linearly divergent triangle diagrams under shifts of the integration variable by a constant (as in the case of the chiral anomaly \cite{Cheng:1984vwu}), do not affect the values of the coefficients in Lagrangians \eqref{Leffff10} and \eqref{Mfi5m}. This is because the corresponding induced contributions enter at a higher order of smallness, namely $\sim 1/M_W^2$, and are therefore negligible within the present approximation.

 Lagrangians \eqref{Leffff10} and \eqref{Mfi5m} fully describe, at leading order, the effective loop-induced interaction of CS bosons with GeV-scale masses with the SM fermions. The interaction of the CS boson with SM fermions obtained in our analysis is similar to the interaction of a $Z^\prime$ boson with SM fermions \cite{Langacker:2008yv,Chiang:2006we,Buras:2012jb}. However, we emphasize that the interaction of the CS boson with same-flavor fermions contains the factor $(\alpha_f +\beta_f \gamma^5)$, whereas the interaction with fermions of different flavors occurs only with left-handed down-type quarks.\vspace{-1.2em}

\section{Conclusions}\vspace{-0.5em}

In this paper, we consider the vector extension of the Standard Model (SM) with Chern-Simons type interaction with a new massive vector boson (Chern-Simons boson) which does not interact directly with fermions of the SM. The interaction of CS bosons with SM fermions occurs solely through loop diagrams.

The effective loop interaction of the CS boson with fermions of different flavors is well defined and free from divergences \cite{Dror:2017ehi,Dror:2017nsg,Borysenkova:2021ydf}. The reason is that the divergent terms of the loop diagrams (containing only the interaction of the CS boson with $W^\pm$ bosons) are proportional to an off-diagonal element of the unit matrix $(V^+V)_{ij}$ ($V$ is the Cabibbo–Kobayashi–Maskawa matrix)   {and therefore cancel as a consequence of the GIM mechanism \cite{Glashow:1970gm,Maiani:2013fpa}. Such a cancellation is a manifestation of a generic feature of flavor-changing neutral-current amplitudes in the SM.}
On the other hand, the effective interaction of the CS boson with same-flavor fermions gives rise to divergences. 
 This problem was investigated in \cite{Borysenkova:2024xno} by considering all possible diagrams in unitary gauge using only the 4-dimensional Lagrangian terms \eqref{Lcs}.
 It was shown that loop divergences could not be eliminated.
In this paper, we consider the renormalizability of the effective loop interaction of the CS boson with fermions of the same flavors (leptons) 
 in the general case of $R_\xi$ gauge with finite values of the gauge parameters $\xi_i$.

We conclude that the loop divergences appearing in the effective interaction of the CS boson with same-flavor fermions cannot be eliminated after summing all relevant diagrams in the $R_\xi$ gauge generated by the interaction Lagrangians \eqref{Lcs} and \eqref{Lcs5}. This remains true both when only the dimension-four interactions of the CS boson with vector fields described by \eqref{Lcs} are included and when the dimension-five CS-boson interactions with the Higgs field, described by \eqref{Lcs5}, are taken into account. The problem is that these divergences cannot be removed via counterterms of the CS boson
interaction with fermions because the initial Lagrangians \eqref{Lcs} and \eqref{Lcs5} do not include these
terms.

Therefore, although the Lagrangian \eqref{Lcs} contains interactions in the form of dimension-four operators for vector fields and may appear renormalizable at first glance   {and therefore consistent with the general arguments of Refs.~\cite{Cornwall:1974km,Liu:2022alx}}, it is in fact nonrenormalizable.   {This result should be understood as a manifestation of the effective-field-theory nature of the gauge-invariant dimension-six operators \eqref{L1} and \eqref{L2}, which give rise to the interactions \eqref{Lcs} after electroweak symmetry breaking.}   We would like to note that from the effective-field-theory perspective, the absence of direct interactions between the CS boson and SM fermions in the initial Lagrangian may be regarded as a matching condition inherited from the ultraviolet completion, reflecting the $U_X(1)$ fermion neutrality. The results obtained in this work show that this low-energy effective theory is not closed under renormalization and therefore requires the inclusion of additional local operators with independent coefficients. These coefficients cannot be determined within the effective theory itself and should be fixed either by matching to a specific ultraviolet completion or by experimental input.

We found that there are only two terms \eqref{divpartL} of the effective loop interaction of the CS bosons with fermions of the same flavors that contain four divergent factors. This conclusion remains valid for calculations performed in both the unitary gauge and the $R_\xi$ gauge.
 Although this paper presents calculations for the interactions of CS bosons with charged leptons, it is clear that the conclusion regarding the impossibility of eliminating divergences in same-flavor fermion interactions also extends to neutrinos and quarks.

 It should be noted that the theory of vector bosons with Chern–Simons–type interactions described by Lagrangians \eqref{L1} and \eqref{L2} was originally developed under the assumption that there are no direct couplings between the CS bosons and SM fermions \cite{Antoniadis:2009ze}. If such interactions were present in the fundamental Lagrangian, they would necessarily appear in the form of dimension-four operators, namely $\bar\ell  \gamma^\mu\ell X_\mu$ and $\bar\ell  \gamma^\mu \gamma^5\ell X_\mu$.  Although one can remove the divergent parameters $\alpha_\ell^{div}$ and $\beta_\ell^{div}$ in \eqref{Ldivpart} by introducing appropriate counterterms, the divergent quantities $\gamma_\ell^{div}$ and $\delta_\ell^{div}$ would nevertheless persist. Consequently, even in this case, the theory would remain non-renormalizable.

Since the initial interaction of the CS bosons with SM fields is given by dimension-six operators \eqref{L1} and \eqref{L2}, we conclude that after the electroweak symmetry breaking the interaction of the CS boson with fermions of the same flavor should be considered within the framework of the effective field theory approach and can be written as \eqref{Leffff}, where the formally infinite coefficients obtained in perturbation theory are replaced by new finite effective couplings.
 It can be seen that this effective interaction becomes multi-parametric, making it analogous in this respect to the effective interaction of axion-like particles with SM particles.

In the leading order, not suppressed by the electroweak scale, we find that the effective loop-induced interaction of SM fermions with a GeV-scale CS bosons is fully described by the Lagrangians \eqref{Leffffsimpli} and \eqref{Mfi5m}. The interaction of the CS boson with SM fermions obtained in our analysis is similar to that of a $Z^\prime$ boson with SM fermions. We note, however, that the interaction with same-flavor fermions includes the factor $(\alpha_f + \beta_f \gamma^5)$, whereas flavor-changing interactions occur only for left-handed down-type quarks. Since the coupling of CS bosons to SM fermions arises through loop corrections, higher-dimensional operators also exist that describe such interactions, see the 5-dimensional operator in \eqref{Leffff} and relation (15) in \cite{Borysenkova:2021ydf}. Another distinctive feature of CS bosons is their three-field interaction of a specific form with SM vector fields, as given by \eqref{Lcs}, and with the Higgs field, as given by \eqref{Lcs5}.

We further assume that, instead of having 22 independent couplings (two for each of the six leptons and five quarks, excluding the top quark), the effective Lagrangian \eqref{Leffffsimpli} in the leading approximation actually contains only 9 independent couplings. 
Specifically, we assume that the effective Lagrangian \eqref{Leffffsimpli}   {can be rewritten in the form \eqref{Leffff10}, which} exhibits approximate universality of the couplings within each fermion family, valid up to a factor of order \((1+\mathcal{O}(m_f^2/M_W^2))\): the same parameters $\alpha_\nu$ apply to all three neutrino species; $\alpha_L$ and $\beta_L$ are approximately the same for all three charged leptons; and $\alpha_U$ and $\beta_U$ are common, within this approximation, to both $u$ and $c$ quarks.  
For down-type quarks, the couplings $\alpha_D$ and $\beta_D$ can be taken to be approximately equal in leading order for the $d$ and $s$ quarks, whereas the $b$ quark generally requires its own independent couplings $\alpha_b$ and $\beta_b$.  
For the interaction of the CS boson with the top quark, one must also take into account contributions described by higher-dimensional operators. 

  {An interesting consequence of our analysis is that, when the imaginary part of the coefficient $C_Y$ associated with the dimension-six operator $\mathcal{L}_1$ \eqref{L1} vanishes, the differences between the effective vector couplings become ultraviolet finite,  at least in the leading approximation. In particular, the finite quantity $\alpha_U-\alpha_D$ \eqref{alphadif} determines the coupling of the CS boson to the neutral component of the isovector vector current of the first-generation quarks through the coupling $\Theta_{W1}$. This result may be useful for future phenomenological studies of the CS boson's hadronic decay channels. It is worth noting that the same coupling $\Theta_{W1}$ also governs the production of the CS boson in heavy-meson decays \cite{Borysenkova:2021ydf}, thereby establishing a connection between heavy-meson searches for the CS boson and its hadronic interactions with light quarks.}

Our results may help guide future searches for a GeV-scale long-lived CS boson in intensity frontier experiments, especially given that the CS boson has already been included in the SHiP experiment proposal \cite{Alekhin:2015byh}.\newpage

 \section*{Acknowledgments}

 The work of V.G. and O.Kh. was supported by the National Research Foundation of Ukraine under project No.2023.03/0149. I.H. acknowledges support by the project 'Search for dark matter and particles beyond the Standard Model' of the Ministry of Education and Science of Ukraine (25BF051-01). The work of O.Kh. was also supported by a grant from Simons Foundation International, SFI-PD-Ukraine-00014573 (PI: L.B.).
 The authors are grateful to Eduard Gorbar and Oleg Ruchayskiy for fruitful discussions and helpful comments.

\appendix

\section{Rules of Feynman diagrams}
\label{AppA}

\setcounter{equation}{0}
\renewcommand{\theequation}{A.\arabic{equation}}

In this paper, we calculate Feynman diagrams using the following rules for vertices with the CS boson, see e.g. \cite{Romao:2012pq,Nagashima:2010jma}. 

\phantom{a}

\begin{minipage}{0.17\textwidth}
    \includegraphics[width=\textwidth]{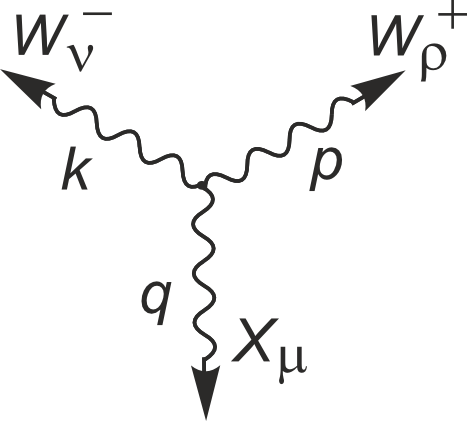}
\end{minipage} %
\begin{minipage}{0.67\textwidth}
    \begin{equation}
    - (c_w p - c_w^* k)_\lambda \epsilon^{\mu\nu\lambda\rho}
    \end{equation}
\end{minipage}

\begin{minipage}{0.17\textwidth}
    \includegraphics[width=\textwidth]{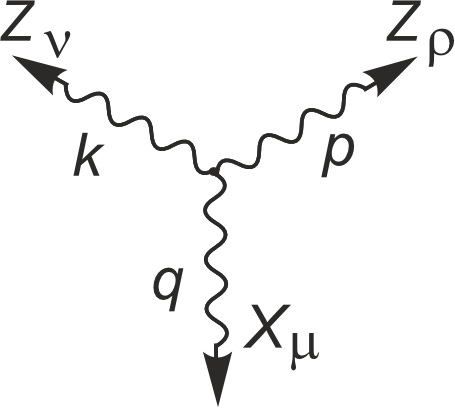}
\end{minipage} %
\begin{minipage}{0.67\textwidth}
    \begin{equation}
  -c_Z\, p_\lambda \epsilon^{\mu\nu\lambda\rho}
    \end{equation}
\end{minipage}

\begin{minipage}{0.17\textwidth}
    \includegraphics[width=\textwidth]{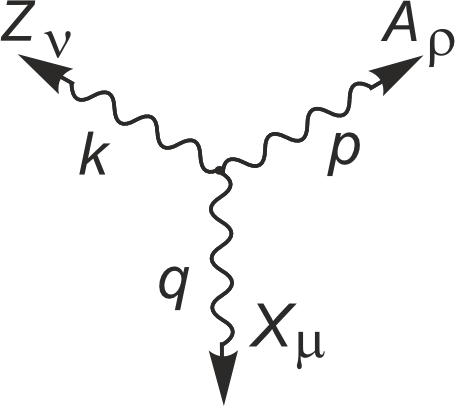}
\end{minipage} %
\begin{minipage}{0.67\textwidth}
    \begin{equation}
    -c_\gamma\, p_\lambda \epsilon^{\mu\nu\lambda\rho}
    \end{equation}
\end{minipage}

\begin{minipage}{0.17\textwidth}
    \includegraphics[width=\textwidth]{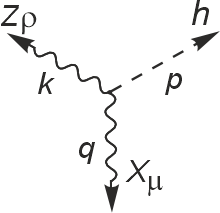}
\end{minipage} %
\begin{minipage}{0.67\textwidth}
    \begin{equation}
    -i \frac{c_{zh}}{v}  p_\nu k_\lambda \epsilon^{\mu\nu\lambda\rho}
    \end{equation}
\end{minipage}

\begin{minipage}{0.17\textwidth}
    \includegraphics[width=\textwidth]{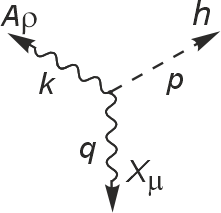}
\end{minipage} %
\begin{minipage}{0.67\textwidth}
    \begin{equation}
    -i \frac{c_{\gamma h}}{v}  p_\nu k_\lambda \epsilon^{\mu\nu\lambda\rho}
    \end{equation}
\end{minipage}

\begin{minipage}{0.17\textwidth}
    \includegraphics[width=\textwidth]{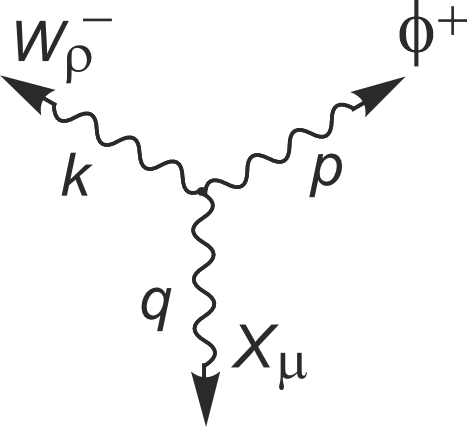}
\end{minipage} %
\begin{minipage}{0.67\textwidth}
    \begin{equation}
   - \frac{2 c_w^{*}}{gv} p_{\nu} k_{\lambda} \epsilon^{\mu\nu\lambda\rho}
    \end{equation}
\end{minipage}

\begin{minipage}{0.17\textwidth}
    \includegraphics[width=\textwidth]{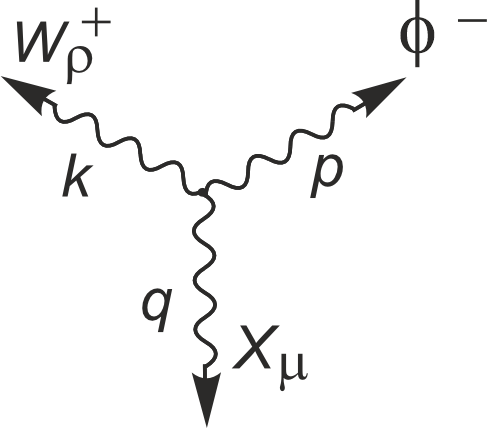}
\end{minipage} %
\begin{minipage}{0.67\textwidth}
    \begin{equation}
    \frac{2 c_w}{gv} p_{\nu} k_{\lambda} \epsilon^{\mu\nu\lambda\rho}
    \end{equation}
\end{minipage}

\begin{minipage}{0.17\textwidth}
    \includegraphics[width=\textwidth]{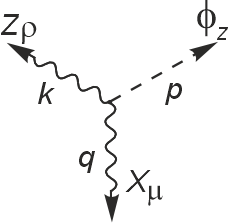}
\end{minipage} %
\begin{minipage}{0.67\textwidth}
    \begin{equation}
    -i \frac{c_z}{v}\frac{2\cos\theta_W}{g }   p_{\nu} k_{\lambda} \epsilon^{\mu\nu\lambda\rho}
    \end{equation}
\end{minipage}

\begin{minipage}{0.17\textwidth}
    \includegraphics[width=\textwidth]{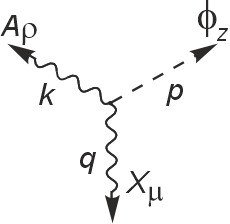}
\end{minipage} %
\begin{minipage}{0.67\textwidth}
    \begin{equation}
   - i\frac{c_\gamma}{v}\frac{2\cos\theta_W}{g }  p_{\nu} k_{\lambda} \epsilon^{\mu\nu\lambda\rho}
    \end{equation}
\end{minipage}

\phantom{vhh}

The propagator in the $R_\xi$ gauge for the fermion fields is given by
\begin{equation}
    G(k) = -i\frac{m+\not{k}}{m^2-k^2}.
\end{equation}

The propagators in the $R_\xi$ gauge for the vector fields $W^\pm$-, $Z$-bosons, and photons are given, respectively, by
\begin{align}
   & D^W_{ij} (k) = \frac{i}{M_W^2 - k^2}\left[g_{ij}+(1-\xi_W)\frac{k_i k_j}{\xi_W M_W^2 - k^2}\right],\\
   & D^Z_{ij} (k) = \frac{i}{M_Z^2 - k^2}\left[g_{ij}+(1-\xi_Z)\frac{k_i k_j}{\xi_Z M_Z^2 - k^2}\right], \\
   & D^A_{ij} (k) = \frac{i}{-k^2}\left[g_{ij}+(1-\xi_A)\frac{k_i k_j}{-k^2}\right].
\end{align}

Propagators in the $R_\xi$ for the scalar fields corresponding to the Goldstone bosons $\phi_Z$, $\phi_{\pm}$, and the neutral Higgs field are given, respectively, by
\begin{equation}
    D^{\phi_{\pm}} (k) = \frac{i}{k^2 - \xi_W M_W^2}, \qquad D^{\phi_Z} (k) = \frac{i}{k^2 - \xi_Z M_Z^2}, \quad D^h (k) = \frac{i}{k^2-m_h^2}.
\end{equation}

Here, $\xi_i$ denotes the corresponding gauge-fixing parameter of the $R_\xi$ gauge.
\newpage

\section{An example of computation of the triangle diagrams}\label{AppB}\vspace{-1.5em}
\setcounter{equation}{0}
\renewcommand{\theequation}{B.\arabic{equation}}
\setcounter{figure}{0}
\renewcommand{\thefigure}{B.\arabic{figure}}

\begin{figure}[h!]\centering
    \includegraphics[width = 0.3\textwidth]{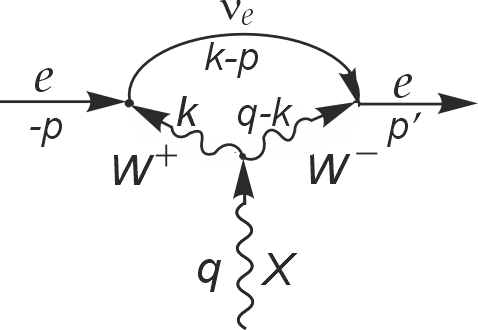}
    \captionsetup{width=0.9\textwidth}
    \caption{Triangle diagram for lepton production via the decay of a CS boson through its interaction with two $W$ bosons.}
    \label{fig:eevWWX}
\end{figure}

The amplitude for the decay of a CS boson into a lepton–antilepton pair in the $R_\xi$ gauge can be written as
\begin{equation}
    M_{fi}^{XWW} = -\frac{g^2}{2} \, \bar{\ell}(p') \hat P_R \mathcal{I}^{\mu} \hat P_L \ell(-p) \epsilon^{\lambda_X}_{\mu}(q),
\end{equation}
where $\mathcal{I}^{ \mu}$ has form
\begin{multline}\label{I}
   \mathcal{I}^{ \mu}
   = \int \dfrac{d^4 k}{(2\pi)^4} \gamma^{\delta} G(k-p) \gamma^\beta D^W_{\delta\nu} (k-q) D^W_{\beta\rho} (k) \left[ c_w k_\lambda - c_w^* (q-k)_\lambda \right] \epsilon^{\mu\nu\lambda\rho} = \\ = 
   \int \dfrac{d^4 k}{(2\pi)^4} \gamma^{\delta} \frac{m_\nu-\not{p}+\not{k}}{m_\nu^2-(k-p)^2} \gamma^\beta \frac{g_{\delta\nu}+(1-\xi_W)\frac{(q-k)_\delta(q-k)_\nu}{\xi_W M_W^2-(k-q)^2}}{M_W^2-(k-q)^2} \frac{g_{\beta\rho}+(1-\xi_W)\frac{k_\beta k_\rho}{\xi_W M_W^2-k^2}}{M_W^2-k^2} \times \\ \times \left[ c_w k_\lambda - c_w^* (q-k)_\lambda \right] \epsilon^{\mu\nu\lambda\rho} = \\ 
    = 
    \gamma^\delta \gamma^\alpha \gamma^\beta \varepsilon^{\mu \nu \lambda \rho}
    \left[c_w g_{\rho \beta} g_{\delta \nu} \int \dfrac{d^4 k}{(2\pi)^4}   
    \dfrac{ (k - p)_\alpha k_\lambda}{(m_\nu^2-(k-p)^2)(M_W^2-(q-k)^2)(M_W^2-k^2)} \right.-\\
    - (1-\xi_W) c_w g_{\rho \beta} \int \dfrac{d^4 k}{(2\pi)^4} \dfrac{ (k - p)_\alpha k_\lambda q_\nu (q-k)_\delta}{(m_\nu^2-(k-p)^2)(M_W^2-(q-k)^2)(M_W^2-k^2)((k-q)^2 - \xi_W M_W^2)} -\\-
    c^*_w g_{\rho \beta} g_{\delta \nu} \int \dfrac{d^4 k}{(2\pi)^4}   
    \dfrac{ (k - p)_\alpha (q-k)_\lambda}{(m_\nu^2-(k-p)^2)(M_W^2-(q-k)^2)(M_W^2-k^2)}+\\ \left.
    +(1-\xi_W) c^*_w g_{\delta \nu} \int \dfrac{d^4 k}{(2\pi)^4} 
    \dfrac{ (k - p)_\alpha q_\lambda k_\beta k_\rho}{(m_\nu^2-(k-p)^2)(M_W^2-(q-k)^2)(M_W^2-k^2)(k^2-\xi_W M_W^2)}\right].
\end{multline}
With the help of this technique, we finally obtain \eqref{XWW}.

\section{An example of integration over the 4-momentum}\label{AppC}
\setcounter{equation}{0}
\renewcommand{\theequation}{C.\arabic{equation}}
\setcounter{figure}{0}
\renewcommand{\thefigure}{C.\arabic{figure}}

Let us calculate $I_{\nu\alpha}(a,\xi_W)$ \eqref{k2integral} using the $\alpha$ (Schwinger) representation, see e.g. \cite{Bogolyubov:1983gp}:
\begin{align}
    &\frac1{m_\nu^2-(k+a)^2-{\rm i}\varepsilon}={\rm i}\int\limits_0^\infty d \tilde\alpha\,
    e^{{\rm i}\tilde\alpha((k+a)^2-m_\nu^2+{\rm i}\varepsilon)},\quad \varepsilon\rightarrow0, \\
    &\frac1{M_W^2-(k-q)^2-{\rm i}\varepsilon}={\rm i}\int\limits_0^\infty d \beta\,
    e^{{\rm i}\beta((k-q)^2-M_W^2+{\rm i}\varepsilon)},\quad \varepsilon\rightarrow0, \\
    &\frac1{\xi_W M_W^2-k^2-{\rm i}\varepsilon}={\rm i}\int\limits_0^\infty d \gamma\,
    e^{{\rm i}\gamma(k^2-\xi_W M_W^2+{\rm i}\varepsilon)},\quad \varepsilon\rightarrow0
\end{align}
where  $\varepsilon\rightarrow0$ (in the following expressions, we will neglect $\epsilon$).

This will allow us to write the expression for $I_{\nu\alpha}(a,\xi_W)$ \eqref{k2integral} in the form
\begin{multline}
    I_{\nu\alpha}(a,\xi_W)=\int \dfrac{d^4 k}{(2\pi)^4}
    \dfrac{k_\nu k_\alpha}{(m_\nu^2-(k+a)^2)(M_W^2-(k-q)^2)(\xi_W M_W^2-k^2)} = \\
    - {\rm i} \int\limits_0^\infty d \tilde\alpha \int\limits_0^\infty d \beta \int\limits_0^\infty d \gamma \int \dfrac{d^4 k}{(2\pi)^4} k_\nu k_\alpha e^{A k^2 + 2B k - M^2},
\end{multline}
where 
\begin{align}
    & A = \tilde\alpha + \beta + \gamma, \quad
     B = \tilde\alpha \, a - \beta \, q \\
    & M^2 = \tilde\alpha \, (m_\nu^2-a^2) + \beta \, (M_W^2 - q^2) + \gamma \, \xi_W M_W^2.
\end{align}
Using relations for integrals over the 4-momentum
\begin{align}
&\int\limits_{-\infty}^\infty d^4k\,e^{{\rm i}(A k^2+2B k)}=\frac
{\pi^2}{{\rm i}}\cdot\frac1{A^2}\,e^{-{\rm i}\frac{B^2}{A}},\label{l21.3a}\\
&\int\limits_{-\infty}^\infty d^4k\,e^{{\rm i}(A k^2+2B k)}k^\nu
=\frac{\pi^2}{{\rm i}}\cdot\frac1{A^2}\,
e^{-{\rm i}\frac{B^2}{A}}\left[-\frac{B^\nu}{A}\right],\label{l21.3b}\\
&\int\limits_{-\infty}^\infty d^4k\, e^{{\rm i}(A k^2+2B k)}k^\nu
k^\mu=\frac{\pi^2}{{\rm i}}\cdot\frac1{A^2}\,
e^{-{\rm i}\frac{B^2}{A}}\left[\frac{2B^\nu B^\mu+{\rm i}A g^{\mu\nu}}{2A^2}\right],\label{l21.3c}\\
& \int\limits_{-\infty}^\infty d^4k\, e^{{\rm i}(A k^2+2B k)}k^\nu k^\mu
k^\lambda=\frac{\pi^2}{{\rm i}}\cdot\frac1{A^2}\,
e^{-{\rm i}\frac{B^2}{A}}\times\nonumber\\
&\hspace{4em} \times\left[-\frac{4B^\mu B^\nu B^\lambda+2{\rm i}A
\left[g^{\mu\nu}B^\lambda+g^{\mu\lambda}B^\nu+g^{\nu\lambda}B^\mu
\right]}{4A^3}\right],\label{l21.3e}
\end{align}
we can write
\begin{equation}
    I_{\nu\alpha}(a,\xi_W) = - {\rm i} \int\limits_0^\infty d \tilde\alpha \int\limits_0^\infty d \beta \int\limits_0^\infty d \gamma \, \frac{\pi^2}{{\rm i}} \frac1{A^2}\,
    e^{-{\rm i}\left(\frac{B^2}{A}+M^2\right)}\left[\frac{2B_\nu B_\alpha+{\rm i}A g_{\nu\alpha}}{2A^2}\right].
\end{equation}
Now, we make a change of integration variables  $\tilde \alpha = xR$, $\beta = yR$, $\gamma = zR$, with an additional condition $x+y+z=1$. The Jacobian  determinant for such a transformation is $I=R^2$, which gives
\begin{align}
    & A = \tilde \alpha + \beta + \gamma = R \, (x+y+z) = R, \\
    & B = R \, (x a-y q), \\
    & M^2 = R(M_W^2 (y+z\xi_W) + x (m_\nu^2-a^2) -y M_X^2),
\end{align}
and
\begin{multline} \label{C13}
    I_{\nu\alpha}(a,\xi_W) = - \frac{\pi^2}{(2\pi)^4} \int \left\{ dx \right\}_3 \int\limits_0^\infty dR \, e^{- {\rm i} R \left((xa-yq)^2 + M_W^2 (y+z\xi_W) + x (m_\nu^2-a^2) -y M_X^2 \right) } \times \\   \times \left[ (xa-yq)_\nu (xa-yq)_\alpha + \frac{i}{2R} g_{\nu\alpha} \right] = \\ - \frac{\pi^2}{(2\pi)^4}\!\! \int\!\! \left\{ dx \right\}_3 \left[ (xa\!-\!yq)_\nu (xa\!-\!yq)_\alpha \int\limits_0^\infty dR \, e^{- {\rm i} R D - \epsilon R } + \frac{i}{2} g_{\nu\alpha} \int\limits_0^\infty \frac{dR}{R} \, e^{- {\rm i} R D - \epsilon R } \right],
\end{multline}
where $\int \left\{ dx \right\}_3$ and $D$ are given by \eqref{Intdx3}, \eqref{DphiW}, and we restored $\epsilon \rightarrow 0$ in the final expression.

The first integral in \eqref{C13} in the limit $\epsilon \rightarrow 0$ is 
\begin{equation}\label{intfirst}
    \int\limits_0^\infty dR \, e^{- {\rm i} R D - \epsilon R } =  \frac{1}{{\rm i}D}.
\end{equation}

To evaluate the second integral, we used the technique of the $\alpha$ (Schwinger) representation and Frullani-based regularization (an analogue of Pauli–Villars regularization) \cite{zuber2012quantum} in the form
    \begin{equation}\label{Frullani1}
        \int\limits_0^\infty \frac{e^{-iR (D-i \varepsilon)}}{R}dR \rightarrow \int\limits_0^\infty \frac{e^{-iR (D-i \varepsilon)}-e^{-iR (D-i \varepsilon)|_{m_{virt}=\Lambda \rightarrow \infty}}}{R}dR,
    \end{equation}
where $m_{virt}$ is the mass of the virtual fermion in the loop. 
So, we get
\begin{multline}
    e^{i R \left[ M_W^2 (y + z\xi_W) - a^2 \, x(1-x) - M_X^2 \, y (1-y) - 2xy \, (aq) + x \, m_\nu^2 \right]}
    \rightarrow\\  e^{i R \left[ M_W^2 (y + z\xi_W) - a^2 \, x(1-x) - M_X^2 \, y (1-y) - 2xy \, (aq) + x \, m_\nu^2 \right]} - e^{-i R \Lambda^2 x}.
\end{multline}

Using the Frullani integral
\begin{equation}\label{Frullani2}
    \int_{0}^{\infty} \frac{dR}{R}\,\bigl(e^{-i a R} - e^{-i b R}\bigr)\,e^{-\epsilon R}
    \;=\;\ln\!\frac{b - i\epsilon}{a - i\epsilon},
\end{equation}
we evaluate the second integral in \eqref{C13}
\begin{equation}\label{intsecond}
    \int_0^\infty \frac{dR}{R} \, e^{- {\rm i} R D - \epsilon R } = \ln \frac{\Lambda^2 x}{D} = \ln \frac{\Lambda^2}{M_W^2} - \ln \frac{D}{M_W^2 x}, \quad  \Lambda\rightarrow \infty.
\end{equation}
Using relations \eqref{intfirst}, \eqref{intsecond}, will finally lead us to the expression  \eqref{k2integral}.

\section{Comparison  with  the case of calculations in the unitary gauge}\label{AppD}
\setcounter{equation}{0}
\renewcommand{\theequation}{D.\arabic{equation}}
\setcounter{figure}{0}
\renewcommand{\thefigure}{D.\arabic{figure}}

It is interesting to note that the calculations in the unitary gauge \cite{Borysenkova:2024xno} also lead to the conclusion that it is impossible to eliminate divergences in the effective loop interaction of the CS boson with leptons of the SM.  Namely, in the unitary gauge, the divergent terms were obtained in the form\footnote{There were some misprints in \cite{Borysenkova:2024xno} that have been corrected in the arXiv version of the paper.}  \cite{Borysenkova:2024xno}
\begin{multline}\label{divpartLunit}
            \sum_{diagrams} M_{fi,div}^{unit} = \Lambda_1  \bar \ell(p^{\prime}) \left[(A+A_5 \gamma^5) \gamma_\nu \gamma_\lambda \gamma_\rho  +\frac{\tilde B_5 \gamma^5}{v}\,q_\nu \gamma_\lambda \gamma_\rho  +\right. \\ \left.
            \frac{C+C_5 \gamma^5}{v^2}\, q_{\nu} p_{\lambda}\gamma_{\rho}+
            \frac{D+D_5 \gamma^5}{v^2}\, q_{\nu} \gamma_{\lambda} \not{q}\gamma_{\rho}+
            \frac{E+E_5 \gamma^5}{v^2}\, q_{\nu} \gamma_{\lambda}\not{p}\gamma_{\rho}\right] \ell(-p)\varepsilon_\mu^{\lambda_X}(q)\,\epsilon^{\mu\nu\lambda\rho},
        \end{multline}
where $A$ and $A_5$ are defined by \eqref{coefA1} and
\begin{multline}\label{coefB}
 \frac{D+D_5 \gamma^5}{v^2}  
 =   \gamma^5 \frac{g}{4 M_W^2} \left[ \frac{c_w g}{2}  + c_z g  t_3^f \left[ t_3^f - 2 q_f \sin^2\theta_W \right] + \ c_{\gamma} q_f e  \cos\theta_W  t_3^f \right] - \\  -\frac{g}{4 M_W^2} \left[ \frac{c_w g}{2} +
     c_z g   \left(  t_3^f \left[ t_3^f - 2 q_f \sin^2\theta_W \right] + 2 q_f^2 \sin^4\theta_W \right) + c_{\gamma} q_f e  \cos\theta_W  \left( t_3^f-2q_f \sin^2\theta_W \right) \right],
\end{multline}\vspace{-1em}
\begin{equation}\label{coefC}
\frac{E+E_5 \gamma^5}{v^2} 
=  i\Theta_W^2 \frac{g^2}{4 M_W^2}(1\!-\!\gamma^5),\,\,
\frac{\tilde B_5 \gamma^5}{v} 
= - \frac{g q_f t_3^f e}{2 \cos\theta_W} \frac{m_f}{M_Z^2} c_\gamma \,  \gamma^5,
\end{equation}\vspace{-1em}
\begin{multline}\label{coefD}
  \frac{C+C_5 \gamma^5}{v^2}
  =  \gamma^5 \frac{g}{3M_W^2} \left[ \frac{\Theta_{W1} g}{2}  + c_z g t_3^f \left[ t_3^f - 2 q_f \sin^2\theta_W \right] + 
      c_{\gamma}  q_f e  \cos\theta_W t_3^f \right] -\\ -  \frac{g}{3M_W^2} \left[ \frac{\Theta_{W1} g}{2}  +  c_z g   \left(  t_3^f \left[ t_3^f - 2 q_f \sin^2\theta_W \right] + 2 q_f^2 \sin^4\theta_W \right) + \vphantom{\frac12} c_{\gamma} q_f e  \cos\theta_W  \left( t_3^f-2q_f \sin^2\theta_W \right) \right].
\end{multline}

At first glance, the structure of divergent terms seems richer and differs from the case of the non-unitary gauge \eqref{divpartL}. But if one uses the equations of motion  
\begin{equation}\label{EOM}
    \bar \ell(p')\! \not{\! p}\,{}' =m_\ell \bar \ell(p'),\qquad -\not{\!p}\,\ell(-p) =m_\ell \ell(-p),
\end{equation}
and relations
\begin{align}
    &     \bar \ell(p') \gamma_{\rho} \not{\!p} \gamma_{\nu}  \ell(-p)=                 
    \bar \ell(p')   2\gamma_{\rho}\, p_{\nu}   \ell(-p) + \bar \ell(p') m_\ell \gamma_{\rho} \gamma_{\nu} \ell(-p),\\
    &     \bar \ell(p') \gamma_{\rho} \not{\!p} \gamma_{\nu} \gamma^5  \ell(-p)=                 
    \bar \ell(p')   2\gamma_{\rho}\, p_{\nu}   \gamma^5 \ell(-p) - \bar \ell(p') m_\ell \gamma_{\rho} \gamma_{\nu} \gamma^5 \ell(-p),\\
    &  \bar \ell(p')  \gamma_{\rho} \not{\!p'} \gamma_{\nu}  \ell(-p)= 
    \bar \ell(p')   2p'_{\rho}\,\gamma_{\nu}    \ell(-p) - \bar \ell(p') m_\ell \gamma_{\rho} \gamma_{\nu} \ell(-p),\\
    & \bar \ell(p')  \gamma_{\rho} \not{\!p'} \gamma_{\nu} \gamma^5  \ell(-p)= 
    \bar \ell(p')   2p'_{\rho}\,\gamma_{\nu}    \gamma^5 \ell(-p) - \bar \ell(p') m_\ell \gamma_{\rho} \gamma_{\nu} \gamma^5 \ell(-p),
    \end{align}
then it can be shown that expression \eqref{divpartLunit} is identical to expression \eqref{divpartL}. Thus, the results of calculations in the unitary and non-unitary gauges are consistent with each other.

\section{Dominant terms mediated by the $XWW$ interaction}\label{AppE}
\setcounter{equation}{0}
\renewcommand{\theequation}{E.\arabic{equation}}
\setcounter{figure}{0}
\renewcommand{\thefigure}{E.\arabic{figure}}

Let us consider the decay of the GeV-scale CS boson into a pair of leptons, $X\rightarrow \bar\ell \ell$, mediated by the $XWW$ interaction in the unitary gauge, see Fig. \ref{fig:eevWWX}. The amplitude for this process has the form
\begin{equation}
    M_{fi}^{XWW} = -\frac{g^2}{2} \, \bar{\ell}(p') \hat P_R \mathcal{I}^{\mu}_\ell \hat P_L \ell(-p) \epsilon^{\lambda_X}_{\mu}(q),
\end{equation}
where $\mathcal{I}^{\mu}_\ell$, after evaluation using the $\alpha$ (Schwinger) representation technique and Frullani-type regularization \eqref{Frullani1}, \eqref{Frullani2}, takes the form
\begin{multline}
\label{E2}
    \mathcal{I}^\mu_\ell =
   \hat \Lambda_{0\ell}^W \left[
   \gamma_{\rho} (\not{\! \mathcal{P}}-\not{\!p})\gamma_{\nu} (\Theta_{W1}(-2\mathcal{P}_{\lambda})\ + c_w^* q_{\lambda}) \right] \epsilon^{\mu\nu\lambda\rho} + \\ +
      \hat \Lambda_{0\ell}^W \frac{q_{\lambda}}{M_W^2}
      \mathcal{P}_{\nu}\gamma_{\rho} \left[c_w (2(p\mathcal{P})+\not{\! q\,}\not{\! \mathcal{P}}-\not{\!q\,}\not{\! p}) - 2\Theta_{W1}  \mathcal{P}^2
      -2i\Theta_W^2  \not{\! p}\not{\! \mathcal{P}}
      \right] \epsilon^{\mu\nu\lambda\rho} + \\ + 
      \hat \Lambda_{1\ell}^W \left[ (-i\Theta_{W1})\gamma_{\rho}\gamma_{\lambda}\gamma_{{\nu}}+ \frac{q_{\lambda}}{M_W^2} \left[
    i \frac{c_w}2  \gamma_{\rho}\not{\!q\,}\gamma_{\nu} -6i\Theta_{W1} \gamma_{\rho}\mathcal{P}_{\nu}+\Theta_W^2 \gamma_{\rho}\not{\!p}\gamma_{\nu}\right] 
    \right] \epsilon^{\mu\nu\lambda\rho},
\end{multline}
where 
\begin{align}
    & \mathcal{P} = xp+yq,\label{Pdefin}\\
    & \hat \Lambda_{0\ell}^W = -\frac{\pi^2}{(2\pi)^4} \int_0^{1}dx\int_0^{1-x}dy\int_0^{\infty}dR\, e^{-iR\,D^W(m_{\nu_\ell})}=i \frac{\pi^2}{(2\pi)^4} \int_0^{1}dx\int_0^{1-x}dy\, \frac{1}{D^W(m_{\nu_\ell})}, \\
     & \hat \Lambda_{1e}^W = -\frac{\pi^2}{(2\pi)^4} \int_0^{1}dx\int_0^{1-x}dy\int_0^{\infty}R^{-1}\,dR\, e^{-iR\,D^W(m_{\nu_\ell})}= -\frac{\pi^2}{(2\pi)^4} \int_0^{1}dx\int_0^{1-x}dy\, \ln\frac{\Lambda^2 x}{D^W(m_{\nu_\ell})}, \\
     & D^W(m_{\nu_\ell})=x m_{\nu_\ell}^2+(1-x)M_W^2-x(1-x)m_\ell^2+y(x+y-1)M_X^2 \label{DWe}.
\end{align}

This expression can be simplified by noting that $M_W^2 \gg M_X^2, m_\ell^2$ and $m_{\nu_\ell}=0$. Using this approximation, we simplify \eqref{DWe} 
\begin{equation}\label{E6}
    D^W = (1-x)M_W^2
\end{equation}
and extract the dominant term of the amplitude, which is not suppressed by the factor $1/M_W$,
\begin{equation}
\label{XWLampl}    
    M_{fi}^{XWW} =  \bar \ell(p') \left[ \frac{g^2}{2} \Theta_{W1} A^W_i \gamma^\mu \gamma^5 \hat P_L \right] \ell(-p) \, \epsilon^{\lambda_X}_\mu(q),
\end{equation}
where $A_\ell^W=6 \hat \Lambda_{1\ell}^W$ and we have taken into account \eqref{e3gamma}.

The operator $\Lambda_{1\ell}^W$ contains a divergent part. As an example, by employing the minimal subtraction (MS) scheme \cite{tHooft:1972tcz,Peskin:1995ev} and removing the divergent term in the integrand,
\begin{equation}\label{MSD}
\ln\frac{\Lambda^2 x}{D^W}
=
\ln\frac{\Lambda^2}{M_W^2}
+
\ln\frac{x M_W^2}{D^W},
\end{equation}
we obtain, using \eqref{E6},
\begin{equation}
A_\ell^{W,\mathrm{MS}}=6\hat \Lambda_{1\ell}^{W,\mathrm{MS}}
=-\frac{6\pi^2}{(2\pi)^4}
\int_0^{1} dx \int_0^{1-x} dy\,
\ln\frac{x M_W^2}{D^W}=1.900 \cdot 10^{-2}.
\end{equation}
This result also holds for neutrinos.

Let us consider the decay of the CS boson into a pair of same-flavor down-type quarks, $X\rightarrow \bar d_n d_n$, in the unitary gauge. The amplitude for this process has the form
\begin{equation}
    M_{fi}^{XWW} = \frac{g^2}{2} \sum_{i=u,c,t} V_{ni}^\dagger V_{in} \bar{d_n} (p') \hat P_R \mathcal{I}^\mu_i \hat P_L d_n (-p) \epsilon^{\lambda_X}_\mu (q),
\end{equation}
where $V_{in}$ are the elements of the CKM matrix. The expression for $\mathcal{I}^\mu_i$ is fully analogous to \eqref{E2}, but it involves different operators, namely $\hat \Lambda^W_0$ and $\hat \Lambda^W_1$:
\begin{align}
    & \hat \Lambda_{0i}^W = -\frac{\pi^2}{(2\pi)^4} \int_0^{1}\!dx\int_0^{1-x}\!dy\int_0^{\infty}dR\, e^{-iR\,D^W(m_{u_i})}=i \frac{\pi^2}{(2\pi)^4} \int_0^{1}\!dx\int_0^{1-x}\!dy\, \frac{1}{D^W(m_{u_i})}, \\
     & \hat \Lambda_{1i}^W = -\frac{\pi^2}{(2\pi)^4} \int_0^{1}dx\int_0^{1-x}dy\int_0^{\infty}R^{-1}\,dR\, e^{-iR\,D^W(m_{u_i})}= -\frac{\pi^2}{(2\pi)^4} \int_0^{1}dx\int_0^{1-x}dy\, \ln\frac{\Lambda^2 x}{D^W(m_{u_i})}, \\
     & D^W(m_{u_i})=x m_{u_i}^2+(1-x)M_W^2-x(1-x)m_{d_n}^2+y(x+y-1)M_X^2 \label{DWWdd}.
\end{align}

To calculate the dominant term for decay into a pair of same-flavor down-type quarks of $n$-generation, we have to take into account that $M_W^2 \gg M_X^2, m_u^2,\, m_c^2$; thus, the expression \eqref{DWWdd} simplifies to
\begin{equation}
\label{DWui}
    D^W(m_{u_i})=
    \left\{\begin{array}{ll}
    D^W_{u,c}= (1-x)M_W^2, & \mbox{for $i=$ u, c}  \\
    D^W_t= x m_t^2+(1-x)M_W^2=M_W^2(1+(t_w-1) x), \,\, t_w=m_t^2/{M_W^2}& \mbox{for $i=$ t}
    \end{array}
    \right.
\end{equation}
The dominant term of the amplitude of decay of the GeV-scale CS boson into a pair of same-flavor down-type quarks is written as
\begin{equation}\label{XWdampl}
    M_{fi}^{XWW} = \bar d_n (p') \left[ \frac{g^2}{2} \Theta_{W1} A_{d_n}^W \gamma^\mu \gamma^5 \hat P_L \right] d_n (-p) \, \epsilon^{\lambda_X}_\mu (q),
\end{equation}
where $A_{d_n}^W=6 \sum\limits_{i=u,c,t} V_{{d_n}i}^\dagger V_{i{d_n}} \hat \Lambda_{1i}^W$ and we have taken into account \eqref{e3gamma}.

Similarly, for decay into a pair of same-flavor up-type quarks ($u$-and $c$-quarks), we use $M_W^2 \gg M_X^2, m_d^2,\, m_s^2,\, m_b^2$, which simplifies \eqref{DWWdd} to \eqref{E6}. 
The dominant term of the amplitude of decay of the CS boson into $\bar u u$ or $\bar c c$ pair  is written as
\begin{equation}\label{XWuampl}
    M_{fi}^{XWW} = \bar u_n (p') \left[ \frac{g^2}{2} \Theta_{W1} A^W_{j} \gamma^\mu \gamma^5 \hat P_L \right] u_n (-p) \, \epsilon^{\lambda_X}_\mu (q),
\end{equation}
where $A_{j}^W=6 \sum\limits_{i=d,s,b} V_{ji}^\dagger V_{ij} \hat \Lambda_{1i}^W$, $j=u,c$ and we have taken into account \eqref{e3gamma}.

Formulas \eqref{XWLampl}, \eqref{XWdampl}, and \eqref{XWuampl}  represent the first term in the expression for the full decay amplitude \eqref{fullamplitude}.

The operator for same-flavor down-type quarks $\Lambda_{1d}^W$ contains a divergent part. As an example, by employing the MS scheme and removing the divergent term in the integrand, see \eqref{MSD}, and using \eqref{DWui},
we obtain, for the case of the decay of the CS boson into a pair of $d$-quarks,
\begin{multline}
    A^{W,MS}_d= 6\sum_{i=u,c,t} V_{di}^\dagger V_{id} \, \hat \Lambda_{1i}^{W,MS} =\\ -\frac{6\pi^2}{(2\pi)^4} \int_0^{1}dx\int_0^{1-x}dy\, \left[ \vert V_{du} \vert^2 \,\ln\frac{x M_W^2}{D^W_u} + \vert V_{dc} \vert^2 \ln\frac{x M_W^2}{D^W_c} + \vert V_{dt} \vert^2 \ln\frac{x M_W^2}{D^W_t} \right] = 1.900 \cdot 10^{-3}.
\end{multline}
As for decay into s- and b-quarks:
\begin{align}
    & A^{W,MS}_s= 6\sum_{i=u,c,t} V_{si}^\dagger V_{is} \, \hat \Lambda_{1i}^{W,MS} = 1.903 \cdot 10^{-2}, \\
    & A^{W,MS}_b= 6 \sum_{i=u,c,t} V_{bi}^\dagger V_{ib} \, \hat \Lambda_{1i}^{W,MS} = 4.206 \cdot 10^{-2}.
\end{align}

Similarly, for the decay of the CS boson into a pair of same-flavor up-type quarks ($\bar u u$- and $\bar c c$-quarks), employing the MS scheme, removing the divergent term in the integrand, and using \eqref{E6}, we obtain
\begin{equation}
  A^{W,MS}_j= 6 \sum_{i=d,s,b} V_{ji}^\dagger V_{ij} \, \hat \Lambda_{1i}^{W,MS} = 1.900 \cdot 10^{-2}, \quad j=u,c.
\end{equation}

As one can see, the coefficient $A^{W,MS}_s$ differs only slightly from the coefficients $A^{W,MS}_d$, $A^{W,MS}_u$, $A^{W,MS}_c$, and $A^{W,MS}_\ell$ when calculated in the leading-order approximation using our MS scheme.

\section{Dominant terms mediated by the $XZZ$ interaction}\label{AppF}

\setcounter{equation}{0}
\renewcommand{\theequation}{F.\arabic{equation}}
\setcounter{figure}{0}
\renewcommand{\thefigure}{F.\arabic{figure}}

Let us consider the decay of the GeV-scale CS boson into a pair of leptons, $X\rightarrow \bar\ell \ell$, mediated by the $XZZ$ interaction in unitary gauge, see Fig.\ref{fig:XZZ_Rx0i}. The amplitude for this process has the form
\begin{equation}
    M_{fi}^{XZZ} = -\frac{g^2}{4\cos^2\theta_W} \, \bar{\ell}(p') \hat{\overline{P}}_Z \mathcal{I}^{\mu,Z}_\ell \hat P_Z \ell(-p) \epsilon^{\lambda_X}_{\mu}(q),
\end{equation}
where $\mathcal{I}^{\mu,Z}_\ell$, after evaluation using the $\alpha$ (Schwinger) representation technique and Frullani-type regularization \eqref{Frullani1}, \eqref{Frullani2}, takes the form
\begin{multline}
\label{F2}
    \mathcal{I}^{\mu,Z}_\ell =
    m_\ell c_z \, \hat\Lambda_{0\ell}^Z \left[ \gamma_\rho \gamma_\nu (q-2\mathcal{P})_{\lambda} + \frac{q_\lambda}{M_Z^2} \mathcal{P}_\nu \not{q} \gamma_\rho \right] \epsilon^{\mu\nu\lambda\rho} + \\ +
   c_z \, \hat\Lambda_{0\ell}^Z \left[
   \gamma_{\rho} (\not{\! \mathcal{P}}-\not{\!p})\gamma_{\nu} (q-2\mathcal{P})_{\lambda} + \frac{q_{\lambda}}{M_Z^2} \mathcal{P}_{\nu}\gamma_{\rho} (-2 \mathcal{P}^2 + 2(p\mathcal{P})+\not{\! q\,}\not{\! \mathcal{P}}-\not{\!q\,}\not{\! p}) \right] \epsilon^{\mu\nu\lambda\rho} + \\ + 
      i c_z \, \hat\Lambda_{1e}^{Z} \left[ -\gamma_{\rho}\gamma_{\lambda}\gamma_{{\nu}}+ \frac{q_{\lambda}}{M_Z^2} \left(
    \frac{1}2  \gamma_{\rho}\not{\!q\,}\gamma_{\nu} - 6 \gamma_{\rho}\mathcal{P}_{\nu} \right) 
    \right] \epsilon^{\mu\nu\lambda\rho}
\end{multline}
where $\mathcal{P}$ is defined in \eqref{Pdefin} and
\begin{align}
    & \hat \Lambda_{0\ell}^Z = -\frac{\pi^2}{(2\pi)^4} \int_0^{1}dx\int_0^{1-x}dy\int_0^{\infty}dR\, e^{-iR\,D^Z(m_\ell)}=i \frac{\pi^2}{(2\pi)^4} \int_0^{1}dx\int_0^{1-x}dy\, \frac{1}{D^Z(m_\ell)}, \\
     & \hat \Lambda_{1\ell}^Z = -\frac{\pi^2}{(2\pi)^4} \int_0^{1}dx\int_0^{1-x}dy\int_0^{\infty}R^{-1}\,dR\, e^{-iR\,D^Z(m_\ell)}= -\frac{\pi^2}{(2\pi)^4} \int_0^{1}dx\int_0^{1-x}dy\, \ln\frac{\Lambda^2 x}{D^Z(m_\ell)}, \\
     & D^Z(m_\ell)=(1-x)M_Z^2+x^2m_\ell^2+y(x+y-1)M_X^2 \label{DZe}.
\end{align}

This expression can be simplified by noting that $M_Z^2 \gg M_X^2, m_\ell^2$. Using this approximation, we simplify \eqref{DZe} 
\begin{equation}\label{F6}
    D^Z = (1-x)M_Z^2
\end{equation}
and extract the dominant term of the amplitude, which is not suppressed by the factor $1/M_W$,
\begin{equation}
\label{XZLampl}
    M_{fi}^{XZZ} = \bar \ell (p') \left[ \frac{g^2}{4 \cos^2\theta_W} c_z A^Z_\ell \gamma^\mu \gamma^5 \hat P_Z^2 \right] \ell (-p) \, \epsilon^{\lambda_X}_\mu (q),
\end{equation}
where $A_\ell^Z=6 \hat \Lambda_{1\ell}^Z$ and we have taken into account \eqref{e3gamma}.

The operator $\Lambda_{1\ell}^Z$ contains a divergent component. For instance, by applying the MS scheme and removing the divergent term from the integrand, see \eqref{MSD}, and using \eqref{F6}, we obtain, for the case of the decay of the CS boson into a pair of charged leptons,
\begin{equation}
  A^{Z,MS}_\ell=6  \hat \Lambda_{1e}^{Z,MS} = -\frac{6\pi^2}{(2\pi)^4} \int_0^{1}dx\int_0^{1-x}dy\,  \,\ln\frac{x M_W^2}{D^Z(m_\ell)}  = 2.38 \cdot 10^{-2}.
\end{equation}
This result also holds for neutrinos.

Let us consider the decay of the CS boson into a pair of same-flavor down-type quarks, $X\rightarrow \bar d_n d_n$, mediated by the $XZZ$ interaction in the unitary gauge. The amplitude for this process has the form
\begin{equation}
    M_{fi}^{XZZ} = -\frac{g^2}{4\cos^2\theta_W} \, \bar{d_n}(p') \hat{\overline{P}}_Z \mathcal{I}^{\mu,Z}_d \hat P_Z d_n(-p) \epsilon^{\lambda_X}_{\mu}(q).
\end{equation}
The expression for $\mathcal{I}^{\mu,Z}_d$ is fully analogous to \eqref{F2}, but it involves different operators, namely $\hat \Lambda^Z_0$ and $\hat \Lambda^Z_1$:
\begin{align}
    & \hat \Lambda_{0n}^Z = -\frac{\pi^2}{(2\pi)^4} \int_0^{1}dx\int_0^{1-x}dy\int_0^{\infty}dR\, e^{-iR\,D^Z(m_{d_n})}=i \frac{\pi^2}{(2\pi)^4} \int_0^{1}dx\int_0^{1-x}dy\, \frac{1}{D^Z(m_{d_n})}, \\
     & \hat \Lambda_{1n}^Z = -\frac{\pi^2}{(2\pi)^4} \int_0^{1}dx\int_0^{1-x}dy\int_0^{\infty}R^{-1}\,dR\, e^{-iR\,D^Z(m_{d_n})}= -\frac{\pi^2}{(2\pi)^4} \int_0^{1}dx\int_0^{1-x}dy\, \ln\frac{\Lambda^2 x}{D^Z(m_{d_n})}, \\
     & D^Z(m_{d_n})= x \, m_{d_n,virt}^2+(1-x)M_Z^2-x(1-x)m_{d_n}^2+y(x+y-1)M_X^2. \label{DZd}
\end{align}

To calculate the dominant terms, we must take into account that $M_Z^2 \gg M_X^2, m_d^2, m_s^2, m_b^2$. In this approximation, the expression \eqref{DZd} simplifies to \eqref{F6}.
The dominant term of the amplitude for the decay of the GeV-scale CS boson into a pair of same-flavor down-type quarks can then be written as
\begin{equation}
\label{XZDampl}
    M_{fi}^{XZZ} = \bar d_n (p') \left[  \frac{g^2}{4 \cos^2\theta_W} c_z \, A^Z_d \, \gamma^\mu \gamma^5 \hat P_Z^2 \right] d_n (-p) \, \epsilon^{\lambda_X}_\mu (q),
\end{equation}
where $A_d^Z=6 \hat \Lambda_{1d}^Z$ and we have taken into account \eqref{e3gamma}.

The same result also applies to the decay of the CS boson into same-flavor up-type quarks ($\bar u u$ and $\bar c c$). After taking into account that $M_Z^2 \gg M_X^2, m_u^2, m_c^2$, the expression for $D^Z(m_{u_n})$ simplifies to \eqref{F6}, and the dominant term of the amplitude becomes
\begin{equation}
\label{XZUampl}
    M_{fi}^{XZZ} = \bar u_n (p') \left[  \frac{g^2}{4 \cos^2\theta_W} c_z \, A^Z_j \, \gamma^\mu \gamma^5 \hat P_Z^2 \right] u_n (-p) \, \epsilon^{\lambda_X}_\mu (q),
\end{equation}
where $A_j^Z=6 \hat \Lambda_{1j}^Z$, $j=u,c$ and we have taken into account \eqref{e3gamma}.

Formulas \eqref{XZLampl}, \eqref{XZDampl}, and \eqref{XZUampl} represent the second term in the expression for the full decay amplitude \eqref{fullamplitude}.

Similarly to the case of the decay of the CS boson into a pair of leptons, we can employ the MS scheme and remove the divergent term in the integrand, see \eqref{MSD}. Using \eqref{F6}, we obtain, for the case of the decay of the CS boson into down-type quarks ($\bar dd$, $\bar s s$, $\bar b b$)
\begin{equation}
  A^{Z,MS}_j=6  \hat \Lambda_{1j}^{Z,MS} = -\frac{6\pi^2}{(2\pi)^4} \int_0^{1}dx\int_0^{1-x}dy\, \left[ \,\ln\frac{x M_W^2}{D^Z(m_{d_j})} \right]  = 2.38 \cdot 10^{-2}, \quad j=d,s,b.
\end{equation}
For the decay of the CS boson into a pair of $u$- or $c$-quarks, using \eqref{F6}, we obtain the same result.

As one can see, the coefficients $A^{Z,MS}_j$ coincide for $j = d, s, b, u, c$, as well as for charged leptons and neutrinos, in the leading-order approximation within our MS scheme.

\section{Dominant terms mediated by the $XZA$ interaction}\label{AppG}
\setcounter{equation}{0}
\renewcommand{\theequation}{G.\arabic{equation}}
\setcounter{figure}{0}
\renewcommand{\thefigure}{G.\arabic{figure}}

Let us consider the decay of the GeV-scale Chern–Simons boson into a pair of leptons, $X \rightarrow \bar\ell \ell$, this time mediated by the channel involving one $Z$ boson and one photon, see Fig.\ref{fig:XZA_Rxi}. The amplitude for this process has the general form
\begin{equation}
    M_{fi}^{XZA} = - \frac{g e q_e}{2\cos\theta_W} c_\gamma \bar{\ell} (p') \left[ \hat{\overline{P}}_Z \mathcal{I}_{1,\ell}^{\mu,ZA} + \mathcal{I}_{2,\ell}^{\mu,ZA}\hat P_Z \right] \ell (-p) \epsilon^{\lambda_X}_\mu (q),
\end{equation}
where $\mathcal{I}_1^{\mu,ZA}$ and $\mathcal{I}_2^{\mu,ZA}$, after evaluation using the $\alpha$ (Schwinger) representation technique and Frullani-type regularization \eqref{Frullani1}, \eqref{Frullani2}, take the form
\begin{multline}\label{I_ZA1}
    \mathcal{I}_{1,\ell}^{\mu,ZA}= 
            - m_\ell \left[ \hat\Lambda_{0,\ell}^{(1\gamma)}\left\{\gamma_{\rho}+\frac{q_{\rho}}{M_Z^2} \,(\not{\!\mathcal{P}} -\not{\!q\,})\right\}\gamma_{\nu}\mathcal{P}_{\lambda}+ \hat\Lambda_{1,\ell}^{(1\gamma)}\frac{i}{2}\frac{q_{\rho}}{M_Z^2}\,\gamma_{\lambda}\gamma_{\nu} \right] \epsilon^{\mu\nu\lambda\rho}+\\
            +
            \hat\Lambda_{0,\ell}^{(1\gamma)}\left\{\gamma_{\rho}(\not{\!p}-\not{\mathcal{\!P}})\gamma_{\nu}+\frac{q_{\rho}}{M_Z^2}
            \left[ (\not{\mathcal{\!P}}\not{\!p} - \mathcal{P}^2)\gamma_{\nu}
            -\not{\!q}\,(\not{\!p}-\not{\mathcal{\!P}})\gamma_{\nu}\right] \right\} \mathcal{P}_{\lambda} \epsilon^{\mu\nu\lambda\rho}+\\
            +
            \hat\Lambda_{1,\ell}^{(1\gamma)}\frac{i}{2}\left\{ -\gamma_{\rho}\gamma_{\lambda}+\frac{q_{\rho}}{M_Z^2}
            \left( \gamma_{\lambda}\not{\!p}+\not{\!q}\, \gamma_{\lambda} - 6\mathcal{P}_{\lambda}\right)\right\}\gamma_{\nu} \epsilon^{\mu\nu\lambda\rho}.
\end{multline}
\begin{multline}\label{I_ZA2}
    \mathcal{I}_{2,\ell}^{\mu,ZA}
    =  - m_\ell \left[ \hat\Lambda_{0,\ell}^{(2\gamma)}\gamma_{\rho} \left\{\gamma_{\nu}(\mathcal{P}_{\lambda}-q_{\lambda})+\frac{q_{\lambda}}{M_Z^2} \not{\!\mathcal{P}}\mathcal{P}_{\nu}\right\} + \hat\Lambda_{1,\ell}^{(2\gamma)}   \frac{i}{2} \gamma_{\rho}\frac{q_{\lambda}}{M_Z^2}\gamma_{\nu}
     \right] \epsilon^{\mu\nu\lambda\rho} + \\
     +
     \hat\Lambda_{0,\ell}^{(2\gamma)}\gamma_{\rho} \left\{(\not{\!p}-\not{\!\mathcal{P}} ) (\mathcal{P}_{\lambda}-q_{\lambda} )\gamma_{\nu}+
     \frac{q_{\lambda}}{M_Z^2} (\not{\!p}\not{\!\mathcal{P}} - \mathcal{P}^2)\mathcal{P}_{\nu} \right\} \epsilon^{\mu\nu\lambda\rho}+\\
     +
     \hat\Lambda_{1,\ell}^{(2\gamma)}\frac{i}{2}\gamma_{\rho}\left\{ -\gamma_{\lambda}\gamma_{\nu}+\frac{q_{\lambda}}{M_Z^2}(\not{\!p} \gamma_{\nu} - 6 \mathcal{P}_{\nu}) \right\} \epsilon^{\mu\nu\lambda\rho},
\end{multline}
where $\mathcal{P}$ is defined in \eqref{Pdefin}, and in the following, we use the notation $i = 1,2$
\begin{align}
    & \hat \Lambda_{0\ell}^{(i\gamma)} = -\frac{\pi^2}{(2\pi)^4} \int_0^{1}dx\int_0^{1-x}dy\int_0^{\infty}dR\, e^{-iR\,D^{(i\gamma)}}=i \frac{\pi^2}{(2\pi)^4} \int_0^{1}dx\int_0^{1-x}dy\, \frac{1}{D^{(i\gamma)}}, \\
     & \hat \Lambda_{1\ell}^{(i\gamma)} = -\frac{\pi^2}{(2\pi)^4} \int_0^{1}dx\int_0^{1-x}dy\int_0^{\infty}R^{-1}\,dR\, e^{-iR\,D^{(i\gamma)}}= -\frac{\pi^2}{(2\pi)^4} \int_0^{1}dx\int_0^{1-x}dy\, \ln\frac{\Lambda^2 x}{D^{(i\gamma)}}, \\
     & D^{(1\gamma)}(m_\ell)=  yM_Z^2+x^2m_{\ell}^2+y(x+y-1)M_X^2, \label{DZAe1} \\
     & D^{(2\gamma)}(m_\ell)= (1-x-y)M_Z^2+x^2 m_{\ell}^2+y(x+y-1)M_X^2. \label{DZAe2}
\end{align}

This expression can be simplified by noting that $M_W^2 \gg M_X^2, m_\ell^2$. Using this approximation, we simplify \eqref{DZAe1}  and \eqref{DZAe2}
\begin{equation}\label{DZAe12sim}
     D^{(1\gamma)}= y M_Z^2, \quad
     D^{(2\gamma)}= (1-x-y)M_Z^2
\end{equation}
and extract the dominant term of the amplitude, which is not suppressed by the factor $1/M_W$,
\begin{equation}
\label{XALampl}
    M_{fi}^{XZA}= \bar{\ell} (p') \left[  \frac{g e q_f}{2 \cos\theta_W} c_\gamma A^\gamma_\ell \gamma^\mu \gamma^5 \hat P_Z \right] \ell (-p) \epsilon^{\lambda_X}_\mu (q),
\end{equation}
where $A_\ell^\gamma=6 (\hat \Lambda_{1\ell}^{(1\gamma)} + \hat \Lambda_{1\ell}^{(2\gamma)}) $ and we have taken into account \eqref{e3gamma}.

The operators $\Lambda_{1\ell}^{(1\gamma)}$ and $\Lambda_{1\ell}^{(2\gamma)}$ contain a divergent component. For instance, by applying the MS scheme and removing the divergent term from the integrand, see \eqref{MSD}, and using \eqref{DZAe12sim}, we obtain, for the case of the decay of the CS boson into a pair of charged leptons,
\begin{equation}
    \hat \Lambda_{1 \, e}^{(1\gamma),MS} = \hat \Lambda_{1 \, e}^{(2\gamma),MS} = 0.80 \cdot 10^{-3}
\end{equation}
and $A^{\gamma,MS}_\ell=9.6\cdot 10^{-3}$. Similarly, one can show that 
\begin{equation}
   A^{\gamma,MS}_f=A^{\gamma,MS}_\ell=9.6\cdot 10^{-3} 
\end{equation}
for all quarks (except for the top quark) and for neutrinos, in the leading-order approximation within our MS scheme.

\newpage

\bibliographystyle{JHEP}
\bibliography{Bibl}

\end{document}